\documentclass[letterpaper, 10 pt, journal, twoside]{IEEEtran}
\pdfoutput=1

\newcommand\Tstrut{\rule{0pt}{2.4ex}}         
\newcommand\Bstrut{\rule[-0.9ex]{0pt}{0pt}}   

\usepackage{textcmds}
\usepackage{bm}
\usepackage{amsmath,amssymb}
\usepackage{enumerate}
\usepackage{graphicx}
\usepackage{tikz}
\usetikzlibrary{shapes.geometric, arrows}
\usepackage{hhline}
\usepackage{longtable}	
\usepackage{multicol}
\usepackage{float}
\usepackage{multirow}
\usepackage{epsf,psfrag,amssymb,amsfonts,color,cite,fancybox}
\usepackage{array}
\usepackage[mathscr]{eucal}
\usepackage{algorithmic}

\usepackage{textcomp}
\usepackage{diagbox}
\usepackage{xcolor}
\usepackage{soul}
\allowdisplaybreaks

\ifCLASSOPTIONcompsoc
\usepackage[caption=false,font=normalsize,labelfon
t=sf,textfont=sf]{subfig}
\else
\usepackage[caption=false,font=footnotesize]{subfi
g}
\fi

\IEEEoverridecommandlockouts

\tikzstyle{startstop} = [rectangle, rounded corners, minimum width=1.2cm, minimum height=0.8cm,text centered, draw=black, fill=white!20]
\tikzstyle{io} = [trapezium, trapezium left angle=70, trapezium right angle=110, minimum width=0cm, minimum height=0cm, text centered, draw=black, fill=white!20]
\tikzstyle{process} = [rectangle, minimum width=0cm, minimum height=0cm, text centered, draw=black, fill=white!20]
\tikzstyle{decision} = [diamond, minimum width=1.2cm, minimum height=0.8cm, text centered, draw=black, fill=white!20]
\tikzstyle{arrow} = [thick,->,>=stealth]

\begin{document}
%
\title{Online PMU-Based Wide-Area Damping Control for Multiple Inter-Area Modes}


%
%

\author{Ilias~Zenelis,~\IEEEmembership{Student Member,~IEEE}, Xiaozhe~Wang,~\IEEEmembership{Member,~IEEE}, and Innocent~Kamwa,~\IEEEmembership{Fellow,~IEEE}
\thanks{This work is supported by the Fonds de Recherche du Qu\'{e}bec - Nature et technologies under Grant FRQ-NT NC-253053 and  FRQ-NT PR-253686.}
\thanks{I. Zenelis and X. Wang are with the Department of Electrical and Computer Engineering, McGill University, Montr\'{e}al, QC  H3A 0G4, Canada (email: ilias.zenelis@mail.mcgill.ca; xw264@cornell.edu).}
\thanks{I. Kamwa is with the Hydro-Qu\'{e}bec/IREQ, Power Systems and Mathematics, Varennes, QC J3X
1S1, Canada (e-mail: kamwa.innocent@ireq.ca).}}

\maketitle

\begin{abstract}
This paper presents a new phasor measurement unit (PMU)-based wide-area damping control (WADC) method to suppress the critical inter-area modes of large-scale power systems. Modal participation factors, estimated by a practically model-free system identification approach, are used to select the most suitable synchronous generators for control through the proposed WADC algorithm. It is shown that multiple inter-area modes can be sufficiently damped by the proposed approach without affecting the rest of the modes, while only a few machines are needed to perform the control. The proposed technique is applied to the IEEE 68-bus and the IEEE 145-bus systems, including the test cases with PMU measurement noise and with missing PMUs. The simulation results clearly demonstrate the good adaptivity of the control strategy subjected to network model changes, its effective damping performance comparing to power system stabilizers (PSSs), and its great potential for near real-time implementation.  
\end{abstract}


\section{Introduction}

\IEEEPARstart{I}{nter-area} oscillations ($0.1-1$ Hz), involving coherent groups of generators swinging relative to each other, commonly appear in modern large interconnected power systems. 
If poorly damped, 
these modes may reduce the transmission capability, restrict the system interconnection, and even lead to partial or total power interruption \cite{Andersson05}. This was the case of the $1996$ Western American Electric Blackout, that has been closely related to the violation of small-signal stability near the equilibrium point. The western inter-area mode, with a frequency of $0.25$ Hz, was lightly damped when the catastrophic event started, and it became unstable (i.e., negatively damped) due to the switching actions that occurred as a consequence of a reactive power deficiency in the Idaho area \cite{Venkatasubramanian04}.

The conventional approach to damp out inter-area oscillations is through power system stabilizers (PSSs) that  provide supplementary control action through the generator excitation systems \cite{Kundur94}. 
Nevertheless, PSSs 
may lack global observability of some inter-area modes \cite{Sallam96}, leading to ineffective damping \cite{Potvin93}, unless 
well-tuned control devices acting in a coordinate manner are employed \cite{Pourbeik98}. 
%
%
In addition, PSSs 
may not be able to adapt to the constant changes in power system operating conditions 
due to generation and load pattern variations \cite{Bose08}, even if multiple input signals are used\cite{Ren00}. 

In recent years, the development of the wide-area measurement system (WAMS) technology
has offered a unique opportunity to overcome the limited modal observability that is inherent to traditional model-based local controllers. In WAMS, phasor measurement units (PMUs) are deployed at strategic locations to record GPS-synchronized power system data at a high speed (i.e., at a $6-60$ Hz sampling rate) \cite{Taylor05},  
which can be used to design wide-area damping controllers for inter-area oscillations. This new family of controllers needs from $4$ to $20$ times smaller gain than local controllers to achieve a similar damping effect \cite{Lefebvre05}, resulting in a reduced control effort. Various methodologies for wide-area damping control (WADC) have been reported in the literature, such as $H_{\infty}$ control \cite{Majumder04}, model predictive control \cite{Iravani13}, nonlinear control \cite{Singh18}, time-delayed control \cite{Patel19}, linear matrix inequalities (LMIs) \cite{Krishnan18}, linear quadratic Gaussian (LQG) control \cite{Preece13}, linear quadratic regulator (LQR)-based optimal control \cite{Bullo14}, model reduction and control inversion \cite{Chakrabortty12}, etc. A comprehensive tutorial on the current wide-area damping control approaches has been presented in \cite{Khargonekar13}. However,  all of these methods depend on accurate information about the network model and its parameters, which, unfortunately, may not be available in real life.  
Besides, most of them require details about the prevailing operating condition and cannot adapt to the continuous changes of the network model. 
Recently, \cite{Zenelis18,Pradhan18} presented methodologies of model-free WADC. However, the proposed methods may not be capable of damping multiple inter-area modes simultaneously. 

In this paper, we seek to propose a novel PMU-based wide-area damping control methodology for   multiple inter-area modes without affecting the other modes. 
Particularly, the hybrid measurement- and model-based approach of estimating the dynamic system state matrix \cite{Bialek17} is applied to extract the modal participation factors, based on which a novel WADC algorithm is designed to damp all the critical inter-area modes. The proposed WADC method is almost purely measurement-based,  requiring only the generator inertias and damping coefficients (i.e., completely independent of the network model). Moreover, the method can realize a full mode decoupling so as to damp all the critical modes without affecting the well-damped local and inter-area modes, using the least possible number of generators. Additionally, the proposed method is computationally efficient and can be implemented in near real-time. Compared to the previous work \cite{Zenelis18}, this paper has made the following additional contributions: 
\begin{itemize}
    \item The WADC algorithm proposed in this paper can damp all the critical inter-area modes simultaneously using the minimum possible number of generators.  
    \item The proposed wide-area damping control scheme is tested and shown to be effective in the cases where PMUs are missing from some generators. It also works well in situations where not all the PMU-equipped generators are capable of conducting the WADC (i.e., only provide data but do not implement control).
    \item The proposed method can quickly adapt to the changes in power network topology or operating condition and still provide the desired damping performance to all the critical inter-area modes, considering the updated modal properties.
\end{itemize}

The remainder of the paper is organized as follows: Section \ref{2} introduces the power system dynamic model and describes the PMU-based mode identification method. Section \ref{3} presents the proposed wide-area damping control algorithm. Simulation study on the IEEE 68-bus and IEEE 145-bus systems is provided in Section \ref{4}. Conclusions and future perspectives are given in Section \ref{5}.

\section{PMU-Based Modal Analysis}\label{2}
In this paper, we study the stability of power systems undergoing small disturbances. 
Specifically, we consider that the load power fluctuates stochastically around the steady-state operating point. Similar to  the approach adopted in \cite{Theresa13,Nwankpa92}, we model the loads as constant impedances experiencing Gaussian variations which are reflected in the diagonal elements of the equivalent admittance matrix $Y$ as seen from the generator internal buses:
\begin{equation}
\label{load_variation}
Y(i,i) = Y_{ii}(1 + \sigma_{i}\xi_i)\angle{\phi_{ii}}, \quad  i\in\{1,...,n \}
\end{equation}
 where $n$ is the number of synchronous generators in the multimachine system, $\xi_{i}$ are mutually independent normal random variables, $\sigma_{i}^2$ describes the strength of load fluctuations and $Y_{ii}\angle{\phi_{ii}}$ is the $\textit{(i,i)}^{th}$ element of the Kron-reduced admittance matrix $Y$ describing the interactions among the generators. We assume that the variations do not affect ${\phi_{ii}}$, indicating a constant load power factor. Such assumption is reasonable in the study of inter-area oscillations where generator dynamics play the most important roles \cite{Chakrabortty12, Nwankpa92}.
\subsection{The Stochastic Power System Model for Inter-Area Modes}
Generator rotor angle and speed 
typically dominate 
the inter-area modes, which means that high-order dynamics do not play a major role on them \cite{Vanfretti10}. 
Hence, the dynamic behavior of inter-area modes can be adequately described by the classical generator model \cite{Kundur94}. By incorporating the load stochasticity of (\ref{load_variation}), the stochastic classical generator model, that is regarded as an equivalent model of aggregated generators \cite{Chow13}, can be represented as \cite{Theresa13}:
\begin{eqnarray}
\dot{\bm{\delta}}&=&\bm{\omega}\label{swingrandom-1}\\
M\dot{\bm{\omega}}&=&\bm{P_m}-\bm{P_e}-{D}\bm{\omega}-E^2G\Sigma\bm{\xi}\label{swingrandom-2}
\end{eqnarray}
with
\begin{equation}
\label{eq:electric_power}
P_{e_{i}} = E_{i}\sum\limits_{j=1}^n E_{j}Y_{ij}\cos(\delta_{i}-\delta_{j}-\phi_{ij}),  \quad  i\in\{1,...,n \}
\end{equation}
where $\bm{\delta}=[\delta_{1},...,\delta_{n}]^T$ is the vector of generator rotor angles, $\bm{\omega}=[\omega_{1},...,\omega_{n}]^T$ is the vector of generator rotor speed deviations from the synchronous speed, $M=\mbox{diag}([M_{1},...,M_{n}])$ is the inertia coefficient matrix, $D=\mbox{diag}([D_1,...,D_n])$ is the damping coefficient matrix, $\bm{P_{m}}=[P_{m_{1}},...,P_{m_{n}}]^T$ is the vector of generator net shaft (mechanical) power inputs, $\bm{P_{e}}=[P_{e_{1}},...,P_{e_{n}}]^T$ is the vector of generator electrical power outputs, $E=\mbox{diag}([E_{1},...,E_{n}])$ is the matrix of generator internal electromotive force (emf) magnitudes, $Y_{ij}\angle{\phi_{ij}}=G_{ij}+jB_{ij}$ is the $\textit{(i,j)}^{th}$ entry of the internal admittance matrix $Y$, $G=\mbox{diag}([G_{11},...,G_{nn}])$, $\Sigma=\mbox{diag}([\sigma_{1},...,\sigma_{n}])$ and $\bm{\xi} = [\xi_{1},...,\xi_{n}]^T$.

The system is characterized by small rotor angle variations due to random load fluctuations. Thus, (\ref{swingrandom-1})--(\ref{swingrandom-2}) can be represented by the following equivalent linearized model near the steady-state equilibrium point $(\bm{\delta_0}, \bm{\omega_0})$: 
\begin{equation}
\label{matrix_form}
\begin{gathered}
\resizebox{1\hsize}{!}{$
\underbrace{
\begin{bmatrix}
    \Delta\dot{\bm{\delta}}\\
    \Delta\dot{\bm{\omega}}
\end{bmatrix}}_{\dot{\bm x}}
= 
\underbrace{
\begin{bmatrix}
  0_{n\times n} & I_{n\times n}\\
  -M^{-1}\frac{\partial {\bm P_{e}}}{\partial {\bm \delta}} & -M^{-1}D
\end{bmatrix}}_{A}
\underbrace{
\begin{bmatrix}
    \Delta\bm{\delta}\\
    \Delta\bm{\omega}
\end{bmatrix}}_{{\bm x}}
+\underbrace{\begin{bmatrix}
0_{n\times n}\\
-M^{-1}E^2G\Sigma
\end{bmatrix}}_{B}
{\bm \xi}$}
\end{gathered}
\end{equation}
where ${\bm x}=[\Delta\bm{\delta},\Delta\bm{\omega}]^T=[\bm{\delta-\delta_0},\bm{\omega-\omega_0}]^T$. Note that $\bm{\delta_0}$ can be obtained from state estimation results, while $\bm{\omega_0}$ is the synchronous speed. 
Specifically, the state vector $\bm{x}$ follows the multivariate Ornstein-Uhlenbeck process\cite{Gardiner09}.   

\subsection{A Hybrid Measurement- and Model-Based Modal Analysis}

We first consider the case in which PMUs 
are available at all the generator terminal buses (optimistic currently, but quite reasonable in the near future) such that the rotor angles $\bm{\delta}$ and speed deviations $\bm{\omega}$ of all the generators can be estimated around steady-state from PMU measurements \cite{Zhou15}, or directly measured \cite{Rahman16}, after which $\bm{x}=[\Delta\bm{\delta},\Delta\bm{\omega}]^T$ can be readily computed. The case where PMUs are missing from some machines will be discussed in Section \ref{ModalanalysismissingPMU}. 
It has been shown in \cite{Wang18} that 
the dynamic state Jacobian matrix $\frac{\partial \bm{P_{e}}}{\partial \bm{\delta}}$ in steady-state operation can be obtained from:
\begin{equation}
\label{dyn_jacobi}
\frac{\partial \bm{P_{e}}}{\partial \bm{\delta}} = MC_{\omega\omega}C_{\delta\delta}^{-1} - DC_{\omega\delta}C_{\delta\delta}^{-1}
\end{equation}
where $C_{\bm x \bm x} = \begin{bmatrix}
    C_{\delta\delta} & C_{\delta\omega}\\
    C_{\omega\delta} & C_{\omega\omega}
\end{bmatrix}$ is the stationary covariance matrix of ${\bm x}$. 
To preserve the flow of the paper, a sketch of  the derivation of (\ref{dyn_jacobi}) is given in Appendix. Following the estimation of $\frac{\partial \bm{P_{e}}}{\partial \bm{\delta}}$, it is straightforward to calculate the dynamic system state matrix $A$ from (\ref{matrix_form}). Apparently and significantly, the estimation of $\frac{\partial \bm{P_{e}}}{\partial \bm{\delta}}$ and $A$ does not require the accurate knowledge of the network model (topology and parameters), and therefore it is 
almost exclusively measurement-based, given that the only assumed knowledge are the machine inertia and damping coefficient matrices $M$ and $D$. 

Although it may be hard to acquire the generator inertia and damping coefficient matrices $M$ and $D$ in large-scale power systems, dynamic equivalencing (or dynamic model reduction) techniques can be applied by power utilities to derive reduced power system models \cite{Chow13, Ma11}. Recent measurement-based model reduction strategies \cite{Salazar11} have allowed the representation of power systems with hundreds of generating units by reduced-order systems with only a few aggregated machines, using wide-area phasor measurements. The inertia constants and damping coefficients for the aggregated equivalent classical generator model can be estimated through PMU measurements \cite{Huang09, Chavan17, Gorbunov20}.

Overall, since PSS\textregistered E software is used in industrial control rooms for wide-area damping control and dynamic state estimation, it can be assumed that $M$ is reasonably known for every single generator of a multimachine power system \cite{Canizares17,King15}. On the other hand, the accurate value of $D$ may not be important for estimating the frequency of inter-area modes. Indeed, according to the Heffron-Phillips approximation of the single-machine infinite bus system \cite{Kundur94}, inexact value of $D$ will 
not greatly affect the modal frequency and eigenstructure (i.e., observability/controllability, residues). 
As a result, a constant $D$ is typically assumed in previous works of WADC 
\cite{Bullo14, Chakrabortty12, Khargonekar13}.

Once $A$ is identified, all the essential modal properties can be extracted. 
The estimated eigenvalues  $\lambda_{{i_+}} = \eta_i+j\omega_i$, $\lambda_{{i_-}} = \eta_i-j\omega_i$, $i\in \{1,...,n\}$ of $A$ form complex conjugate pairs, 
which define the electromechanical oscillatory modes in terms of oscillation frequency $f_i=\frac{\omega_i}{2\pi}$, damping ratio $\zeta_i=\frac{-\eta_i}{\mid{\lambda_{i_\pm}}\mid}$ and settling time $t_{S_i}=\frac{4}{\mid{\eta_i}\mid}$\cite{Ogata10}. 
An inter-area mode $i$ is called critical if it is poorly damped,  
i.e., its damping ratio $\zeta_i(\%)$ is less than $10\%$,  and/or its settling time $t_{S_i}$ is more than $10s$. These thresholds are recommended in the NERC reliability standards \cite{NERC18}, incorporated into the grid codes and operation guidelines of many utilities \cite{Paserba96}, and widely used in the literature (e.g., \cite{Singh13}).
Furthermore, the right eigenvector matrix $\Phi=[\phi_1,...,\phi_{2n}]$ and the left eigenvector matrix $\Psi=[\psi_1^T,...,\psi_{2n}^T]^T$ of $A$ can be estimated.  
The participation matrix $P=[P_1,...,P_{2n}]$, where
\begin{equation}
\label{part_factor}
P_i=[P_{1,i},...,P_{2n,i}]^T=[\vert{\phi_{1,i}\psi_{i,1}}\vert,...,\vert{\phi_{2n,i}\psi_{i,2n}}\vert]^T
\end{equation}
can also be obtained from the estimated right and left eigenvectors. Particularly, the participation factor $\vert{\phi_{j,i}\psi_{i,j}}\vert, j\in\{1,...,2n\}$ describes the relative participation of the $j^{th}$ state variable in the $i^{th}$ mode, thus being effective on pinpointing key locations to conduct WADC measures \cite{Iravani13, Pal05}.  


\subsection{Modal Analysis with Missing PMUs}\label{ModalanalysismissingPMU}
In real-life applications, missing PMUs cases are inevitable due to device malfunctions, economical considerations, etc. Therefore, we consider a more general situation where PMUs are available only at a set of generators $ {\bm G_P}\subseteq \{1,...,n \}$.   
The proposed mode identification method is still able to estimate the submatrix $(\frac{\partial \bm{P_{e}}}{\partial \bm{\delta}})_P=\frac{\partial \bm{P_{e}}}{\partial \bm{\delta}}[i,j]$, 
where $[i,j]$ denotes the element in row $i$ and column $j$, and 
$i\in {\bm G_P}$, $j\in {\bm G_P}$. Specifically, 
\begin{equation}
\label{dyn_jacobi_missingPMU}
(\frac{\partial \bm{P_{e}}}{\partial \bm{\delta}})_P = M_PC_{\omega\omega_P}C_{\delta\delta_P}^{-1} - D_PC_{\omega\delta_P}C_{\delta\delta_P}^{-1}
\end{equation}
where $M_P=M[i,i]$, $D_P=D[i,i]$, $C_{\omega\omega_P}=C_{\omega\omega}[i,j]$, $C_{\delta\delta_P}=C_{\delta\delta}[i,j]$,  $C_{\omega\delta_P}=C_{\omega\delta}[i,j]$, 
and $i,j\in {\bm G_P}$. Moreover, the corresponding submatrix of $A$ can be constructed:
\begin{eqnarray}
\label{modstatematrix_missingPMU}
A_P&=&\left[\begin{array}{cc}{{0_{m\times m}}}&{I_{m\times m}}\\-M_P^{-1}(\frac{\partial \bm{P_{e}}}{\partial \bm{\delta}})_P&-M_P^{-1}D_P\end{array}\right]
\end{eqnarray}
where $m$ is the cardinality of ${\bm G_P}$.

\section{Wide-Area Damping Control Implementation}\label{3}

\subsection{Linear State Feedback Control}

Once the dynamic system state matrix $A_P$ is estimated using the technique developed in Section \ref{2} (i.e., from (\ref{dyn_jacobi_missingPMU})--(\ref{modstatematrix_missingPMU})), we intend to use the linear state feedback control method \cite{Bullo14} to 
achieve a desired damping behavior for the closed-loop power system described by the following state-space equation: 
\begin{eqnarray}
\label{eq:feedback}
\nonumber\dot{\bm{x}} &=& A_P\bm{x}+B_{c}\bm{u} +B_P\bm{\xi}\\
\bm{u} &=& K\bm{x}
\end{eqnarray}
where the state vector $\bm{x}=[\Delta\bm{\delta},\Delta\bm{\omega}]^T$ is computed from the  measurements of PMUs at the generators in set ${\bm G_P}$, and $B_P=[0,-M_P^{-1}E_P^2G_{GP}\Sigma_P]^T$
with $E_P=E[i,i]$, $G_{GP}=G[i,i]$ and $\Sigma_P=\Sigma[i,i]$; $i\in {\bm G_P}$. The control matrix $B_c$ and the gain matrix $K$ are the design parameters.  

Particularly, the control matrix $B_{c}$ is used to describe which generators 
are selected to perform the wide-area damping control. These generators are termed as \qq{WADC-selected}. Hence, $B_{c}$ can be written as:
\begin{equation}\label{control_matrix}
B_{c} = \begin{bmatrix}
    {0} & 0\\
    0 & B_{c\omega}
\end{bmatrix}  
= \begin{bmatrix}
    0 & 0\\
    0 & \mbox{diag}([b_{c\omega_1},...,b_{c\omega_m}])
\end{bmatrix} 
\end{equation}
where:
\begin{equation}
\label{b_elements}
b_{c\omega_i} =
\begin{cases}
  1, \quad \mbox{if generator $i$ is WADC-selected}\\ 
  0, \quad \mbox{otherwise} 
\end{cases}
\end{equation}
Note that the upper left submatrix of $B_c$ is $\bm{0}$, indicating that the control signals $\bm{u}$ work only on the rotor speed deviation vector $\bm{\omega}$ but not on the rotor angle vector ${\bm \delta}$, since the latter would be physically infeasible.

Ideally, we hope that all the PMU-equipped generating units in ${\bm G_P}$ can conduct the WADC through, for example, supplementary control of generator excitation, turbine governors, high voltage direct current
(HVDC) links, and FACTS devices \cite{Iravani13}. However, not all the generators have the ability to do the control in real life due to several reasons (e.g., outdated infrastructure). Therefore, the WADC-selected machines may only come from ${\bm G_A}$, 
a subset of $ {\bm G_P}$ containing all the PMU-equipped generators that are capable of performing control. 

The goal of the WADC algorithm to be designed is to minimize the number of WADC-selected machines while achieving the desired damping performance for all the critical inter-area modes. The achievement of such goal hinges on the design of the gain matrix $K$. It will be shown in Section \ref{designK} that a gain matrix $K$ is designed to simultaneously damp all the critical modes without affecting the other modes.  
\subsection{Design of the Gain Matrix $K$}\label{designK}
Inspired by the model-based method introduced in \cite{Far09}, we propose a  gain matrix $K$ as the following: 
\begin{equation}
\label{gain_matrix}
K = -B_c\sum_{k\in C_r} \sigma_{d_{k}}[\phi_{k_+},\phi_{k_-}][\psi_{k_+}^T,\psi_{k_-}^T]^T
\end{equation}
where $C_r$ is the set including  all the critical inter-area modes that need to be damped simultaneously.  
For each critical mode $k$ defined by the eigenvalues 
$\lambda_{{k_+}} = \eta_k+j\omega_k$, $\lambda_{{k_-}} = \eta_k-j\omega_k$, 
$\phi_{k_+}$ and $\psi_{k_+}$ are the right and the left eigenvectors of $\lambda_{k_+}$; $\phi_{k_-}$ and $\psi_{k_-}$ are the right and the left eigenvectors of $\lambda_{k_-}$. $\sigma_{d_{k}}>0$ is a chosen damping coefficient for each mode $k$ that can be physically determined by the operating reserve of synchronous generators. 


To see how the proposed linear state feedback control method modulates the eigenvalues of $A_P$ and thus the oscillation modes, we re-write the model (\ref{eq:feedback}) as:
\begin{eqnarray}
\label{feed_rewritten}
\nonumber
\dot{\bm{x}} &=&
(A_P+B_cK)\bm{x}+B_P\bm{\xi}\\\nonumber&=&
(A_P + \Delta A_P)\bm{x}+B_P\bm{\xi}\\&=&A_{cl}\bm{x}+B_P\bm{\xi}
\end{eqnarray}
where $A_{cl} = A_P + \Delta A_P$ is the closed-loop dynamic system state matrix. By the matrix perturbation theory \cite{Stewart90}, the eigenvalues $[\tilde{\lambda}_1,...,\tilde{\lambda}_{2m}]$ of $A_{cl}$ can be represented as the  eigenvalues of $A_P$ plus a perturbation resulting from $\Delta A_P$:
\begin{equation}\label{closed_eig}
\begin{split}
\tilde{\lambda}_i&=\lambda_i+\frac{\psi_{i}\Delta A_P\phi_{i}}{\psi_{i}\phi_{i}}, \quad i\in\{1,...,2m\}
\end{split}
\end{equation}
Using the orthogonality of eigenvectors:
\begin{equation}
\label{eq:orthogonality}
\psi_j\phi_i =
\begin{cases}
  1, \quad \mbox{if } i = j \\
  0, \quad \mbox{if } i \neq j
\end{cases}
\end{equation}
where a vector normalization has been applied, 
$\tilde{\lambda}_i$, $i\in\{1,...,2m\}$ in (\ref{closed_eig}) can be further simplified as shown below: 
%
\begin{align}\label{closed_eig_v2}
\tilde{\lambda}_i&=\lambda_i+\psi_{i}\Delta A_P\phi_{i}\nonumber\\&=
\lambda_i+\psi_{i}B_c K\phi_{i}\nonumber\\&=
\lambda_i+\psi_{i}B_c\left(-B_c\sum_{k\in C_r} \sigma_{d_{k}}[\phi_{k_+},\phi_{k_-}][\psi_{k_+}^T,\psi_{k_-}^T]^T \right)\phi_{i}\nonumber\\&=
\lambda_i-\psi_{i}B_c\left(\sum_{k\in C_r} \sigma_{d_{k}}[\phi_{k_+}\psi_{k_+}+\phi_{k_-}\psi_{k_-}] \right)\phi_{i}\nonumber\\&=
\lambda_i-\sum_{k\in C_r} \sigma_{d_{k}}\psi_{i}B_c[\phi_{k_+}\psi_{k_+}\phi_{i}+\phi_{k_-}\psi_{k_-}\phi_{i}]\nonumber\\&=
\begin{cases}
  \lambda_i-\sigma_{d_k} \psi_{i}B_c\phi_{i}, \quad \mbox{if } i\in\{k_+, k_-\}, k\in C_r\\
  \lambda_i, \quad \mbox{otherwise}
\end{cases}
\end{align}
Therefore, the eigenvalues $\lambda_k=\eta_k\pm j\omega_k$ of the critical inter-area mode $k\in C_r$ are shifted in the complex plane, 
while the other well-damped modes are not affected by the proposed control method. 
The product $\psi_{i}B_c\phi_{i}$ is typically a complex number. 
Consequently, to ensure an improved damping performance for the closed-loop system (\ref{feed_rewritten}), it is crucial to select the best generator combination among the available ones to perform the WADC. 
In this paper, the optimum generator combination, denoted by ${\bm O_G}$, is selected from the generators that have the largest participation factors with respect to (w.r.t.) the critical modes. 
More details are to be discussed in Section \ref{WADCalgorithm}. 



\subsection{The WADC Algorithm for Multiple Inter-Area Modes}\label{WADCalgorithm}
Let $\tilde{\lambda}_k=\eta_{cl_k}\pm j\omega_{cl_k}$ be the closed-loop eigenvalues of the critical inter-area mode $k\in C_r$. Then the objective of the proposed WADC method is to ensure that all the critical modes meet the following requirements: 
\begin{enumerate}[{(i)}]
\item The closed-loop damping ratio $\zeta_{cl_k}=\frac{-\eta_{cl_k}}{\mid{\tilde{\lambda}_k}\mid}$ satisfies:
 \begin{equation}\label{specification1}
    \zeta_{cl_k}\geq0.1, \quad k\in C_r
    \end{equation}
\item The closed-loop settling time $t_{Scl_k}=\frac{4}{\mid{\eta_{cl_k}}\mid}$ satisfies:
 \begin{equation}\label{specification2}
    t_{Scl_k}\leq10s, \quad k\in C_r
    \end{equation}
\item Minimum possible number of generators should be used to reduce communication cost \cite{Bullo14}.
\end{enumerate}


In view of the above, the following index is used to describe the damping performance of the closed-loop system:
\begin{equation}\label{perf}
    J = \frac{1}{\left\lVert S_\zeta\right\rVert}\sum_{k\in C_r} \zeta_{cl_k} - \frac{1}{\left\lVert S_{t_S}\right\rVert}\sum_{k\in C_r} t_{Scl_k}
\end{equation}
where $\left\lVert S_\zeta\right\rVert=\sqrt{\sum_{k\in C_r} \zeta_{cl_k}^2}$, $\left\lVert S_{t_S}\right\rVert=\sqrt{\sum_{k\in C_r} t_{Scl_k}^2}$, i.e., a normalization has been applied. The objective of the proposed WADC method is to maximize $J$ while minimizing $n_m$, i.e., the number of WADC-selected generators. 

\begin{figure}[!h]
\begin{center}
\begin{tikzpicture}[node distance=0.9cm]
\tikzstyle{every node}=[font=\footnotesize]
\node (xi) [process, xshift=2.3cm] {\textbf{1)} Calculate the state vector $\bm{x}=[\Delta\bm{\delta},\Delta\bm{\omega}]^T$ from PMU measurements};
\node (jacobi) [process, below of=xi, text width=6.5cm, yshift=-0.2cm] {\textbf{2)} Estimate the dynamic state Jacobian matrix $(\frac{\partial \bm{P_{e}}}{\partial \bm{\delta}})_P$ by (\ref{dyn_jacobi_missingPMU}) and construct the state matrix $A_P$ by (\ref{modstatematrix_missingPMU})}; 
\node (crit) [process, below of=jacobi, text width=7.4cm, yshift=-0.5cm] {\textbf{3)} Determine the $n(C_r)$ critical inter-area modes by calculating the eigenvalues $\lambda_{{i}} = \eta_i\pm j\omega_i$, $i\in\{1,...,m\}$  
of $A_P$, their damping ratios $\zeta_i$, and their settling times $t_{S_i}$};
\node (part) [process, below of=crit, text width=7.8cm, yshift=-0.63cm] {\textbf{4)}
For each critical mode $k \in C_r$, sort the generators in ${\bm G_P}$ from the one having the largest participation factor to the one having the smallest participation factor, and, thus, create an ordered list}; 
\node (1mach) [io, below of=part, yshift=-0.28cm] {Set $n_m=1$};
\node (liste) [process, below of=1mach, text width=4cm, yshift=-0.28cm]{\textbf{5)} 
Take from each ordered list the first $n_m$ generators and put them in a list ${\bm G_E}$ without duplicates};
\node (GEcheck) [decision, below of =liste, yshift=-1.6cm] {${\bm G_E}\cap{\bm G_A}=\O$ {\large ?}};
\node (GEYES) [io, right of=GEcheck, text width=1.7cm, xshift=2.25cm] {$n_m=n_m+1$};
\node (GENO) [process, below of=GEcheck, yshift=-1.42cm,text width=3.6cm] {\textbf{6)} Find all the combinations of $n_m$ elements in ${\bm G_E}\cap{\bm G_A}$}; 
\node (damp) [process, below of=GENO, text width=5.4cm, yshift=-0.62cm]{\textbf{7)} Compute the control matrix $B_{c}$ by  (\ref{control_matrix}), the gain matrix $K$ by (\ref{gain_matrix}), the damping ratios $\zeta_{cl_k}$ and settling times $t_{Scl_k}, k\in C_r$, and the index $J$ by (\ref{perf}) for each combination};
\node (listc) [process, below of=damp, yshift=-0.65cm, text width=4.5cm] {\textbf{8)} Create a list ${\bm C_{el}}$ containing all the combinations that satisfy (\ref{specification1}) and (\ref{specification2})};
\node (CELcheck) [decision, below of =listc, yshift=-1.05cm] {$\bm C_{el}=\O$ {\large ?}};
\node (CELNO) [process, below of=CELcheck, text width = 5.5cm, yshift=-1.18cm] {\textbf{9)} Select from ${\bm C_{el}}$ the generator combination ${\bm O_G}$ that maximizes the index $J$ and send $\bm{u} = K\bm{x}$ to the $n_m$ generators in ${\bm O_G}$};
\draw [arrow] (xi) -- (jacobi);
\draw [arrow] (jacobi) -- (crit);
\draw [arrow] (crit) -- (part);
\draw [arrow] (part) -- (1mach);
\draw [arrow] (1mach) -- (liste);
\draw [arrow] (liste) -- (GEcheck);
\draw [arrow] (GEcheck) -- node[anchor=south, xshift=-0.1cm, yshift=0.1cm] {Yes} (GEYES);
\draw [arrow] (GEcheck) -- node[anchor=east, xshift=-0.1cm, yshift=0.1cm] {No} (GENO);
\draw [arrow] (GEYES) |- (liste);
\draw [arrow] (GENO) -- (damp);
\draw [arrow] (damp) -- (listc);
\draw [arrow] (listc) -- (CELcheck);
\draw [arrow] (CELcheck) -| node[anchor=east, xshift=-0.7cm, yshift=0.3cm] {Yes} (GEYES);
\draw [arrow] (CELcheck) -- node[anchor=east, xshift=-0.1cm, yshift=0.1cm] {No} (CELNO);
\end{tikzpicture}
\caption{Flowchart of the proposed WADC algorithm.}
\label{chart}
\end{center}
\end{figure}
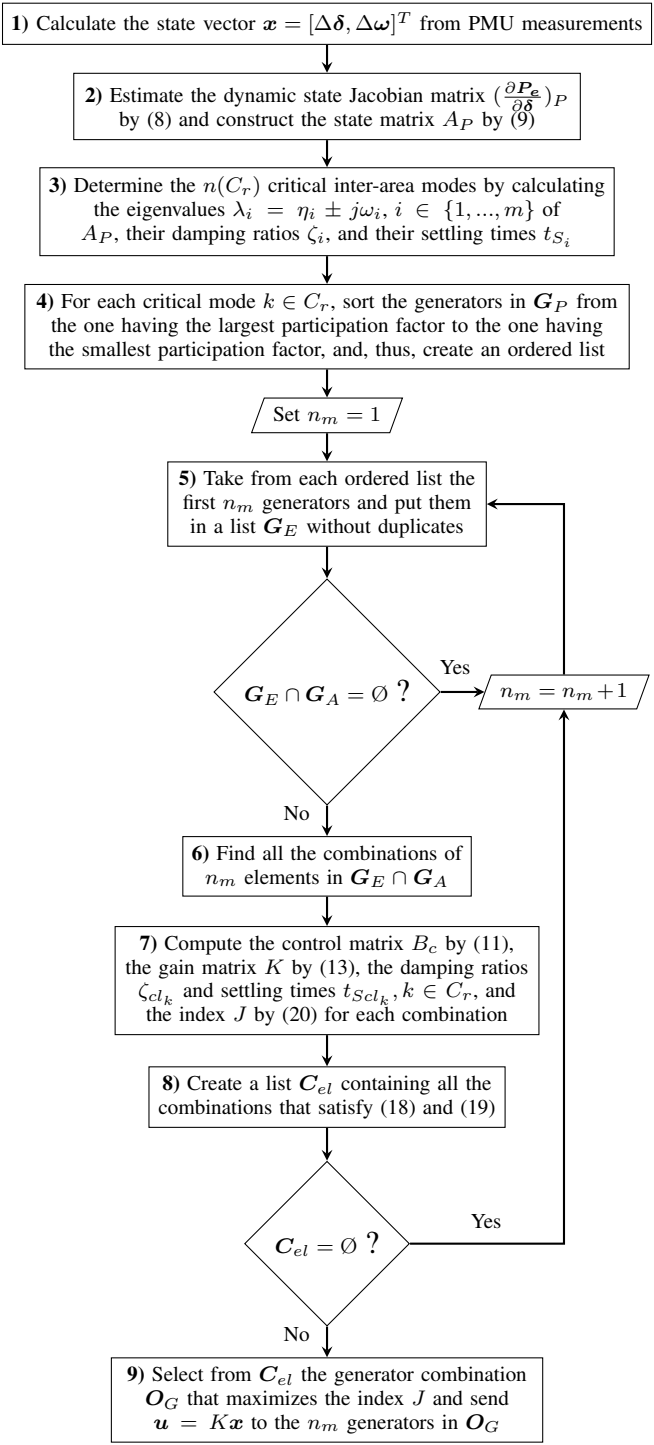

Assume that $\sigma_{d_{k}}, k\in C_r$ have already been decided by the operating reserve of synchronous generators.
The proposed WADC algorithm, aiming to damp all the critical modes simultaneously using the least possible number of generators, is described in Fig. \ref{chart}.

For the proposed WADC algorithm the following assumptions have been made:
\begin{itemize}
    \item ${\bm G_A}\neq\O$, i.e., at least one generator is available for control. 
    \item $\sigma_{d_{k}}, k\in C_r$ are large enough to ensure that ${\bm O_G}$ can be found. The worst case scenario is that ${\bm O_G}$ is achieved for $n_m=m$.
\end{itemize}

It should be noted that the  $10\%$ damping ratio threshold can also guarantee a maximum value of percent overshoot (PO), 
which characterizes the difference between the peak value and the steady-state value of an inter-area oscillation 
so that the oscillation amplitude remains bounded 
\cite{Ogata10}.

Moreover, in the proposed algorithm,  we intend to minimize the number of WADC-selected generators so as to reduce the potential communication cost as suggested in \cite{Bose08, Bullo14, Far09}. However, an alternative WADC formulation may be designed to describe the WADC control effort using the index
\begin{equation}\label{alter_J}
    J_C=\sum_{k\in C_r} \sigma_{d_k}
\end{equation}
such that the total control effort  distributed among all generators in ${\bm G_A}$ is minimized. To this end, after Step $3$, starting from small initial values, the coefficients $\sigma_{d_k}$ for all $k\in C_r$ can be gradually increased until all modes in $C_r$ become well-damped or until the selected limit for $\sigma_{d_{k}}$ is reached.  

\section{Results and Discussion}\label{4}

\begin{figure}[!t]
\centering
\includegraphics[width=2.5in ,keepaspectratio=true,angle=0]{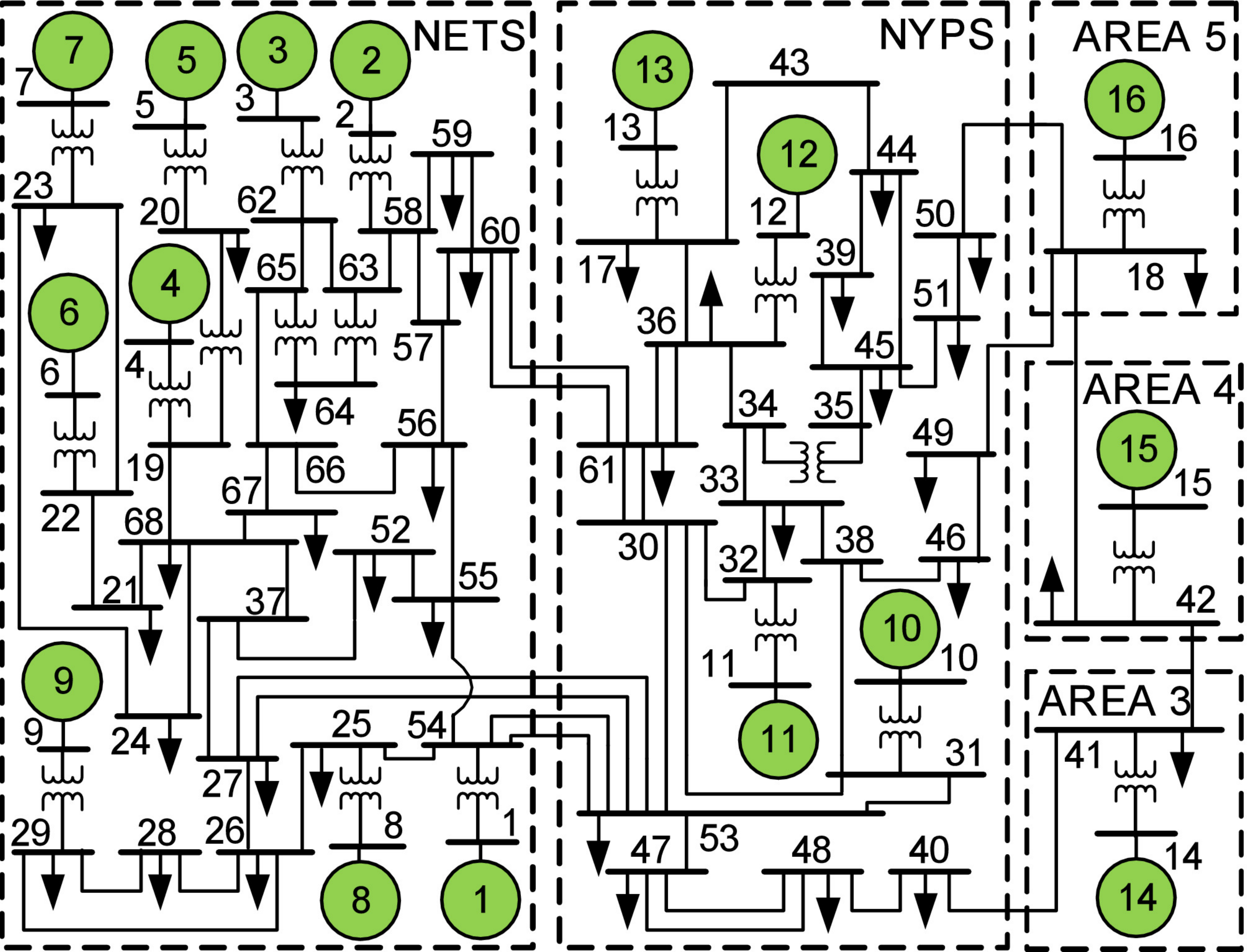}
\caption{68-bus, 16-generator, 5-area benchmark system \cite{Pal05}.}\label{NETS-NYPS_1}
\end{figure}

The primary benchmark used for simulations is the IEEE 68-bus, 16-generator, 5-area power system (Fig. \ref{NETS-NYPS_1}). The system data were adopted from \cite{Pal05}, with some of the parameters modified to meet the requirements of PSAT Toolbox \cite{Milano05} that was employed to carry out the numerical experiments. Although the applied PMU-based mode identification technique (see Section \ref{2}) and the proposed WADC method (see Section \ref{3}) were developed using the classical generator model,
simulations have been performed using the 6th-order generator model 
to show the feasibility of the methods in practical applications. 
Machine $G16$ is the reference generator. Moreover, $G1-G12$ are controlled by automatic voltage regulators (AVRs), and $G9$ is also equipped with a PSS (as defined in \cite{Pal05}). 
A sampling rate of $60$ Hz, which lies in the typical PMU reporting rate range ($6-60$ Hz) \cite{Khargonekar13}, has been used for the time-domain simulations.
All the network parameters are available in: https://github.com/zenili/Wide-Area-Damping-Control-2019. 

\subsection{Validation of the WADC Algorithm in Damping Multiple Inter-Area Modes Under PMU Measurement Noise}\label{subsection_validation}

In this section, we intend to validate the PMU-based mode identification method described in  
Section \ref{2} and, more importantly, the WADC algorithm for multiple mode damping proposed in Section \ref{3}. 
For illustration purposes, we first assume that PMUs are installed at all the generator terminal buses and that all generators are able to perform the control. 
 The intensities $\sigma_i$ due to random load variations, i.e., the diagonal entries of $\Sigma$ in (\ref{swingrandom-2}),  are set to be $20$.

\begin{table}[!t]
\centering
\caption{Actual and Estimated Inter-Area Mode Frequencies}\label{freq_open}
\begin{tabular}{c c c c}
\hhline{====}
\hline
Mode&Actual $f$ (Hz) & Estimated $f$ (Hz) &Error (\%) \Tstrut\Bstrut\\
  \hline
  1&0.418&0.408&2.392\Tstrut\\
  2&0.632&0.580&8.228\\
  3&0.773&0.714&7.633\Bstrut\\
\hhline{====}
  \end{tabular}
 \end{table}

 \begin{table}[!t]
\centering
  \caption{Actual and Estimated Inter-Area Mode Damping Ratios}\label{damping_open}
  \begin{tabular}{c c c c}
\hhline{====}
\hline
Mode&Actual $\zeta$ (\%) & Estimated $\zeta$ (\%) &Error (\%) \Tstrut\Bstrut\\
  \hline
  1&2.871&2.933&2.160\Tstrut\\
  2&1.491&1.610&7.981\\
  3&1.715&1.852&7.988\Bstrut\\
\hhline{====}
  \end{tabular}
 \end{table}
 
 \begin{table}[!t]
\centering
  \caption{Actual and Estimated Inter-Area Mode Settling  Times}\label{settling_open}
    \begin{tabular}{c c c c}
\hhline{====}
\hline
Mode&Actual $t_S$ (s) & Estimated $t_S$ (s) &Error (\%) \Tstrut\Bstrut\\
  \hline
  1&53.065&53.116&0.096\Tstrut\\
  2&67.512&68.142&0.933\\
  3&48.028&48.162&0.279\Bstrut\\
\hhline{====}
  \end{tabular}
 \end{table}

Particularly, $180s$ PMU data is used to estimate the covariance matrix $C_{\bm x \bm x}$, the dynamic state Jacobian matrix $\frac{\partial \bm{P_{e}}}{\partial \bm{\delta}}$ by (\ref{dyn_jacobi}), and the state matrix $A$ by (\ref{matrix_form}).
A zero-mean Gaussian measurement noise with standard deviation of $10^{-3}$ for angles and $10^{-6}$ for speeds is added to the emulated PMU measurements \cite{Anagnostou18}.
Tables  \ref{freq_open}-\ref{settling_open} present the estimated modal properties of all the inter-area modes including the frequency, the damping ratio and the settling time, in comparison with the actual ones.  
As can be seen, the actual modal properties are accurately estimated despite the presence of PMU measurement noise. 
In addition, all the inter-area modes are poorly damped, i.e., their damping ratios are less than $10\%$ and their settling times are more than $10s$. Hence, $C_r=\{1,2,3 \}$, and $n(C_r)=3$. 

\begin{figure}[!t]
\centering
\includegraphics[width=3in ,keepaspectratio=true,angle=0]{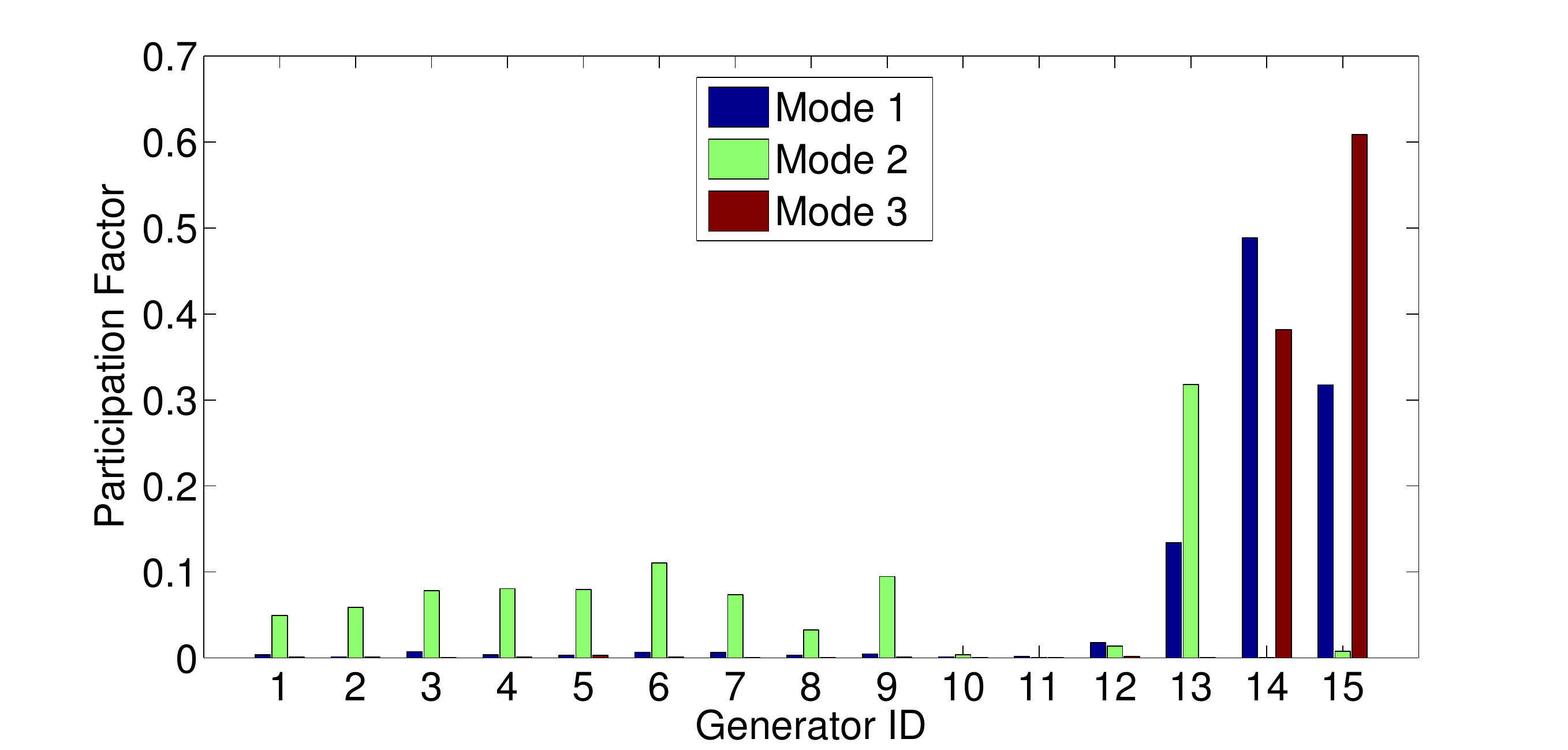}
\caption{Estimated modal participation factors of the critical inter-area modes.}\label{partc}
\end{figure}

Therefore, we apply the proposed WADC algorithm to damp all the critical modes simultaneously using the minimum possible number of generators, without affecting the other modes. The same damping coefficients $\sigma_{d_{k}}=2, k\in C_r$ are chosen for all the inter-area modes.
In Step $4$, the estimated modal participation factors of the critical modes  (Fig. \ref{partc}) are used to create an ordered generator list for each mode.  
Starting from $n_m=1$, ${\bm G_E}=\{13,14,15\}$ contains the generators with the largest participation factors w.r.t all the modes. After calculating the required quantities (i.e., $B_c, K, \zeta_{cl_k},t_{Scl_k},k\in C_r$ and $J$) for each combination of one element in ${\bm G_E}$ (Steps $6-7$), it turns out that when the control is performed by $\{G15\}$, 
$J$ reaches its maximum value. In this case, the damping performances of all the critical modes are presented in the first row of Table \ref{dampingperformance-1}. As can be observed, the damping ratio and the settling time of Mode $2$ do not satisfy (\ref{specification1}) and (\ref{specification2}). Thus ${\bm C}_{el}=\O$ and $n_m$ has to be increased by $1$. When $n_m=2$,  ${\bm G_E}=\{6,13,14,15\}$ according to Fig. \ref{partc}, and all the combinations of $2$ generators from ${\bm G_E}$ are tried. As it turns out, when the control is conducted by $\{G13,G15\}$, the maximum $J$ is achieved. Nonetheless, the damping ratio and the settling time of Mode $2$ still do not meet the requirements (\ref{specification1}) and (\ref{specification2}), as shown in the second row of Table \ref{dampingperformance-1}. Finally, $n_m$ is increased to $3$, and ${\bm G_E}=\{5,6,9,13,14,15\}$. The maximum $J$ is reached when the control is performed by ${\bm O_G}=\{G6,G13,G15\}$. The final damping performances of all the critical inter-area modes are shown in the last row of Table  \ref{dampingperformance-1}. The damping ratios of all the  modes are greater than $10\%$ and their settling times are less than $10s$. 

\begin{table}[!t]
\centering
  \caption{Comparison Between the Open-Loop and the Closed-Loop Properties of the Critical Modes}\label{closed_loop_results}
  \setlength{\tabcolsep}{2pt}
  \begin{tabular}{c c c c c c c}
\hhline{=======}
\hline
$n_m$&Mode&Est. $\zeta_{ol}$ (\%)&$\zeta_{cl}$ (\%)&Est. $t_{Sol}$ (s) &$t_{Scl}$ (s) &Combination\Tstrut\Bstrut\\
  \hline
  1&1&2.933&15.495&53.116&9.755
&\{G15\}\Tstrut\\\
   &2&1.610&1.799&68.142&60.971&\\
   &3&1.852&15.595&48.162&5.909&\Bstrut\\\hline
  2&1&2.933&21.323&53.116&7.024&\{G13, G15\}\Tstrut\\\
   &2&1.610&9.971&68.142&11.055&\\
   &3&1.852&15.503&48.162&5.938&\Bstrut\\\hline
  3&1&2.933&21.445&53.116&6.999&\textbf{\{G6, G13, G15\}}\Tstrut\\\
   &2&1.610&13.211&68.142&8.328&\\
   &3&1.852&15.438&48.162&5.963&\Bstrut\\\hline
\hhline{=======}
\vspace{-5pt} 
  \end{tabular}\label{dampingperformance-1}
\raggedright{Note: \qq{Est.} stands for \qq{Estimated}, \qq{$_{ol}$} for \qq{open-loop}, and \qq{$_{cl}$} for \qq{closed-loop}. }
 \end{table}
 
\begin{figure}[!t]
\centering
\includegraphics[width=2.2in ,keepaspectratio=true,angle=0]{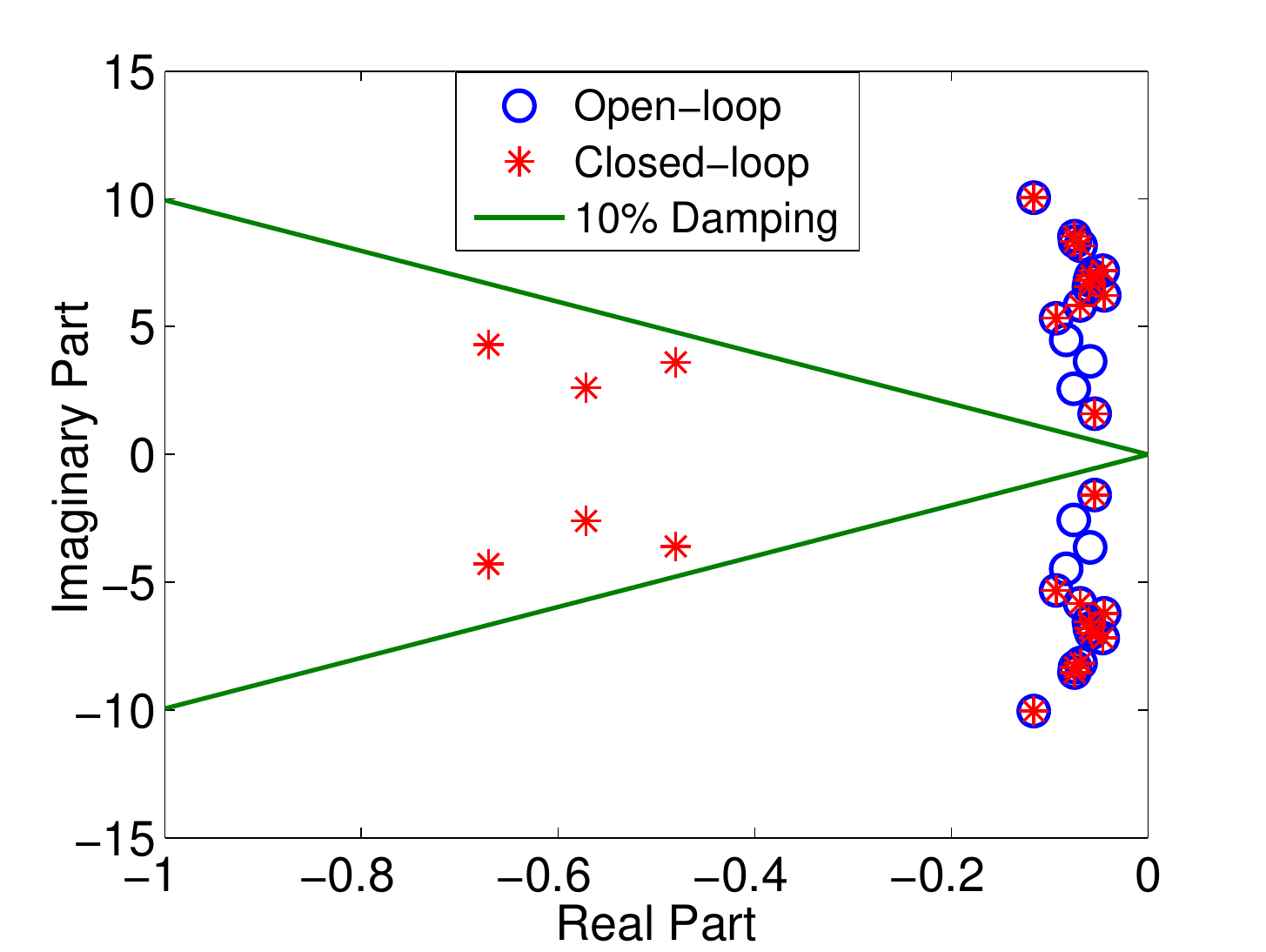}
\caption{Open-loop and closed-loop eigenvalues of all the modes.}\label{response2}
\end{figure}

To demonstrate the impact of the proposed WADC method on all the system's modes,
a comparison between the open-loop and the closed-loop eigenvalues is shown in Fig. \ref{response2}. The eigenpairs corresponding to the three critical inter-area modes have migrated to the left and entered the $10\%$ damping ratio region (defined by the $10\%$ damping lines), as expected. Moreover, the other eigenvalues are not affected by the proposed method, indicating a full decoupling between modes. 

It is worth mentioning that the approximate CPU time consumed by the WADC algorithm to find ${\bm O_G}$ and provide the WADC-selected generators with the wide-area signals ${\bm u}$ (Steps $3-9$) is $0.03s$, including the trials of all the combinations in Step $7$. The short execution time shows that the proposed WADC algorithm can run in  near real-time, while at the same time being robust against PMU measurement noise.

Clearly, there is always a trade-off between the values of $\sigma_{d_{k}}, k\in C_r$, which depend on the system's operating reserve, and the number of the required generators to perform the control. In fact, $n_m$ decreases as $\sigma_{d_{k}}$ increases so that the desired damping goal can be achieved with less WADC-selected generators. However, large $\sigma_{d_{k}}$ indicates a large operating reserve, which may lead to an inefficient utilization of the power generation; small $\sigma_{d_{k}}$, on the other hand, may be insufficient for the desired damping performance even when all the available generators participate in the WADC. 
It is therefore important for a power system operator to adopt a reasonable strategy of selecting $\sigma_{d_{k}}$. 

Lastly, the performance of the WADC algorithm under time delay has been studied. According to \cite{Majumder04, Cheng14}, 
communication time delay 
may negatively affect the performance of the proposed wide-area damping control method. Particularly, if a single time delay $\tau_d$ is accounted for, the closed-loop equation for the state feedback-controlled power system can be written as \cite{Singh09}:
\begin{align}\label{eq:feedback_td}
\nonumber\dot{\bm{x}}(t) &= A_P\bm{x}(t)+B_{c}\bm{u}(t-\tau_d) +B_P\bm{\xi}\\
\bm{u}(t-\tau_d) &= K\bm{x}(t-\tau_d)
\end{align}
The characteristic equation that corresponds to the time-delayed system (\ref{eq:feedback_td}) is now
\begin{equation}\label{transen}
    \det(-sI+A_P+B_cKe^{-\tau_d s})=0
\end{equation}
The eigenvalues of the closed-loop system are the roots of (\ref{transen}). An approximate solution to this delay-differential equation can be computed using a discretization of its partial differential equation (PDE) representation \cite{Bellen00}. In our simulation study, it has been shown that if $\tau_d$ is smaller than $10ms$, which represents a typical value of regional latency in power systems where generators have a wide spatial distribution \cite{Paul18}, its impact on the inter-area mode damping ratios is negligible. Otherwise, it becomes significant and a compensation approach  \cite{Ren00, Iravani13, Cheng14} should be considered to ensure the desired damping performance and the stability of the power system.

\subsection{WADC Application in the Case of  Missing PMUs}
In this numerical study, we attempt to show that the proposed WADC method still works effectively in a more realistic setup where 
PMUs are missing from some generators and/or some units are unable to implement the WADC via the feedback signals $\bm{u}$. To do so, we assume that 
the PMU at $G10$ is missing,  
i.e., ${\bm G_P}=\{1,2,...,9,11,...,15\}$ and $G13$ cannot be used for the WADC, i.e., ${\bm G_A}={\bm G_P}\setminus\{13\}$.

Table \ref{miss_pmu_results} summarizes the estimated frequencies, damping ratios and settling times for all the critical inter-area modes. 
As can be seen, the modal estimation is highly accurate, with the estimated quantities remaining close to the actual ones shown in Tables  \ref{freq_open}-\ref{settling_open}, although the PMU at $G10$ is missing. 
It is worth noting that 
the state variables of $G10$ have small participation in the excitation of the critical modes (see Fig. \ref{partc}), and thus the missing PMU does not greatly affect the accuracy of the mode identification. Nevertheless, a loss of PMU in larger participants (i.e., $G13,G14$ and $G15$) could have a more considerable impact on the estimated results.    

\begin{table}[!t]
\centering
  \caption{Estimated Frequencies, Damping Ratios and Settling Times of the Critical Modes with a Missing PMU}\label{miss_pmu_results}
  \setlength{\tabcolsep}{1.6pt} 
  \begin{tabular}{c c c c c c c}
\hhline{=======}
Mode&Est. $f$ (Hz)&Error (\%)&Est. $\zeta (\%)$ &Error (\%)&Est. $t_{S}$ (s)&Error (\%)\Tstrut\Bstrut\\
  \hline
  1&0.411&1.675&2.814&1.985&55.072&3.782\Tstrut\\
  2&0.583&7.753&1.653&10.865&66.024&2.204\\
  3&0.714&7.633&1.853&8.047&48.147&0.248\Bstrut\\\hline
\hhline{=======}
  \end{tabular}
 \end{table}	

The results of the WADC algorithm application  with $\sigma_{d_{k}}=2, k\in C_r$ are shown in Table \ref{closed_loop_results_inc}. As can be observed, the unavailability of $G13$ for control has a notable effect on the resulting ${\bm O_G}$. Five generators, i.e., $n_m=5$, have to be used to perform the proposed WADC  
instead of three (see Section \ref{4}-A), leading to an increased control effort and  a growing demand for online spinning reserve. Indeed, $G13$ constitutes a substantial part of the proposed WADC algorithm due to its big participation in Mode $2$, and thus three extra generators (i.e., $G3$, $G4$ and $G9$) are needed to compensate for its contribution. 
\begin{table}[!t]
\centering
  \caption{Comparison Between the Open-Loop and the Closed-Loop Properties of the Critical Modes with a Missing PMU}\label{closed_loop_results_inc}
  \setlength{\tabcolsep}{1.5pt}
  \begin{tabular}{c c c c c c}
\hhline{======}
\hline
Mode&Est. $\zeta_{ol}$ (\%)&$\zeta_{cl}$ (\%)&Est. $t_{Sol}$ (s) &$t_{Scl}$ (s) &${\bm O_G}$\Tstrut\Bstrut\\
  \hline
  1&2.814&15.411&55.072&9.715&\textbf{\{G3, G4, G6, G9, G15\}}\Tstrut\\
  2&1.653&12.780&66.024&8.616&\\
  3&1.853&15.413&48.147&5.954&\Bstrut\\\hline
\hhline{======}
\end{tabular}
 \end{table}

\subsection{WADC Adaptivity Assessment and Comparison with PSS}

\subsubsection{Operation under Altered Modal Properties}\label{intention}

One of the most important features of the proposed WADC algorithm is its quick adaptivity to the real-time changes of the network model. 
Such changes may excite different oscillation modes and/or 
lead to different modal participation factors. 
Conventional PSS control, 
depending on offline parameter tuning, may not operate effectively
in real-world situations in which a change in the network model affects the oscillation modes and their modal properties. 
In contrast, the proposed WADC algorithm is able to realize a full decoupling between modes and conduct a real-time adaptive control to damp all the critical modes simultaneously using a small number of generators.  

To show that, we have executed dynamic simulations on a modified version of the 68-bus system, with slightly increased generator damping coefficients, assuming that the system is in a more damped situation. Particularly, we assume that the line between Bus $58$ and Bus $63$ has an outage at $t=1s$,  
making the system slightly perturbed from the new steady state. 
The modal characteristics of inter-area Mode $2$ have been evaluated for both the nominal condition (i.e., where the system operates around steady state under the influence of random load fluctuations), and the topology change condition under three different control scenarios:
\begin{enumerate}[(a)]
    \item No control
    \item PSS control at $G1-G12$
    \item WADC at two generators (i.e., $n_m=2$)
\end{enumerate}
The proposed WADC algorithm with $\sigma_{d_{2}}=2$, targeting only the critical Mode $2$ for illustration purposes, 
is used in the last scenario,
while it is assumed that all generators are PMU-equipped and can perform the WADC. Besides, all generators are modelled by the 6th-order model, with $\sigma_i=5$ (see (\ref{swingrandom-2})).  

\begin{figure}[!t]
\centering
\includegraphics[width=3in ,keepaspectratio=true,angle=0]{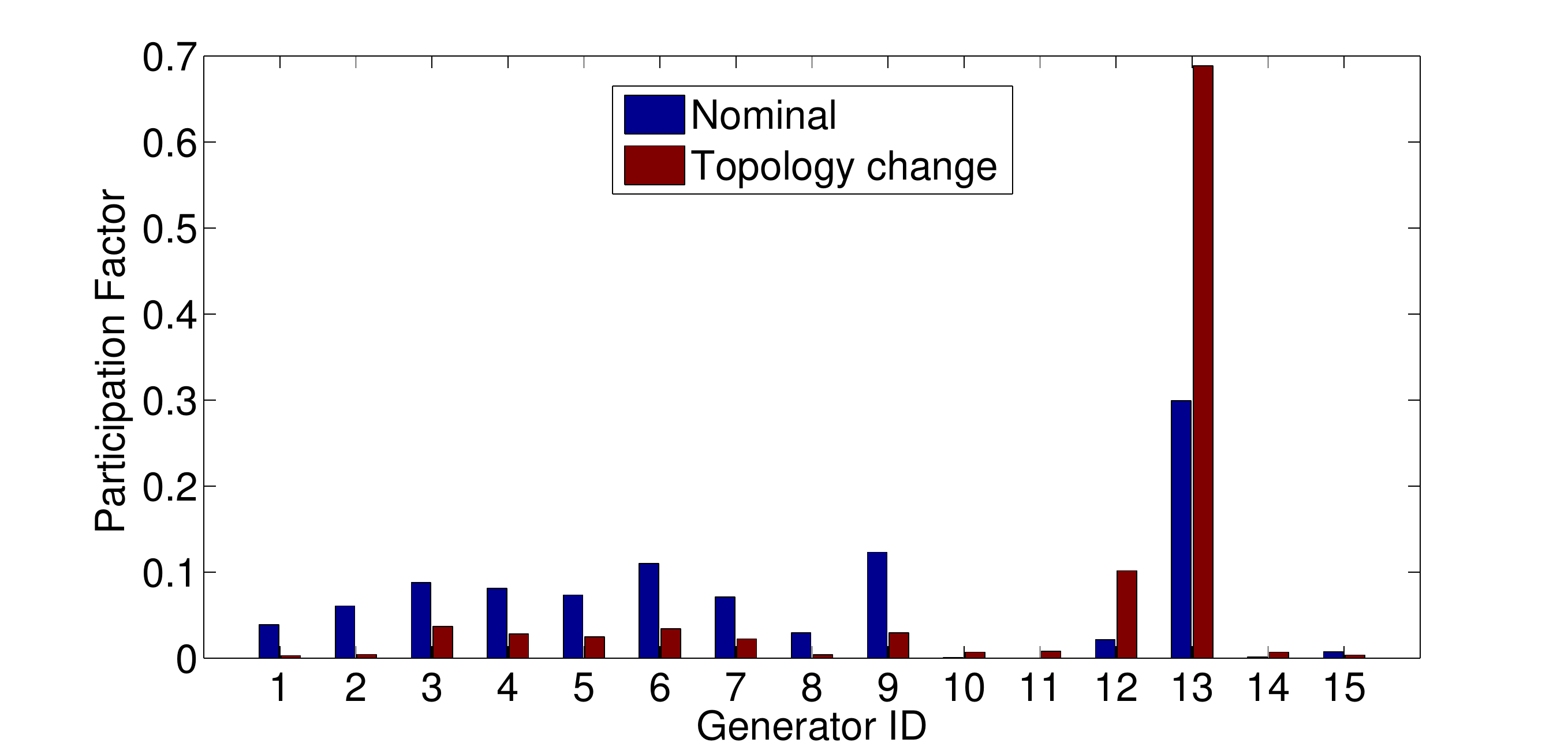}
\caption{Estimated modal participation factors of Mode 2 under different operating conditions.}\label{part2}
\end{figure}

Particularly, the inter-area mode properties are estimated using PMU measurements from the interval $[100s, 200s]$, when the system has sufficiently settled down to the new steady state such that linear modal analysis can be performed. 
The estimated participation factors of Mode 2 under scenario (a) are given in Fig. \ref{part2}. 
It can be seen that the participation factor ranking subject to the topology change differs from the nominal one. With the exception of $G13$ that remains the most important participant in Mode $2$, modal participation is distributed differently among the synchronous machines. As a result, the generator allocation for the proposed wide-area damping control scheme should be adapted accordingly. 

A comparison between the damping performance of Mode $2$ under the different control scenarios (a)-(c) with and without the line outage is presented in Table \ref{condition_results}. In the nominal condition, 
PSSs are well-tuned to provide enough damping to Mode $2$. 
In the topology change condition, PSSs are incapable of meeting the $10\%$ damping ratio and the $10s$ settling time thresholds for Mode $2$, because the PSS parameters cannot be adaptively modified.
An offline parameter re-tuning procedure is required. 
In contrast, the proposed WADC algorithm keeps its entire functionality despite the transmission line outage, thanks to its online adaptive nature. Indeed, the proposed WADC method produces even better results in terms of damping performance for Mode $2$, as the participation of $G13$ in Mode $2$ is almost doubled after the outage of line $58-63$.
 
\begin{table}[!t]
\centering
  \caption{Estimated Properties of Mode 2 under Different  Operating Conditions and Control Scenarios}\label{condition_results}
\setlength{\tabcolsep}{4.4pt} 
  \begin{tabular}{c c c c c}
\hhline{=====}
\multirow{3}{4em}{Operating Condition} & \multirow{3}{4em}{Control Scenario} &&&\\
&&Est. $f$ (Hz) &Est. $\zeta (\%)$ &Est. $t_{S}$ (s)\\
&&&&\\
  \hline
Nominal&No control&0.587&8.220&13.157\Tstrut\\
          &PSS control&0.545&14.119&8.187\\
          &WADC (G13, G9)&0.595&19.132&5.487\Bstrut\\\hline
Topology&No control&0.538&3.708&31.898\Tstrut\\
           change&PSS control&0.576&7.151&15.428\\
           &WADC (G13, G12)&0.528&27.076&4.290\Bstrut\\
\hhline{=====}
  \end{tabular}\label{prevspost}
 \end{table}

\begin{figure}[!t]
\centering
\subfloat[]{\includegraphics[width=3in]{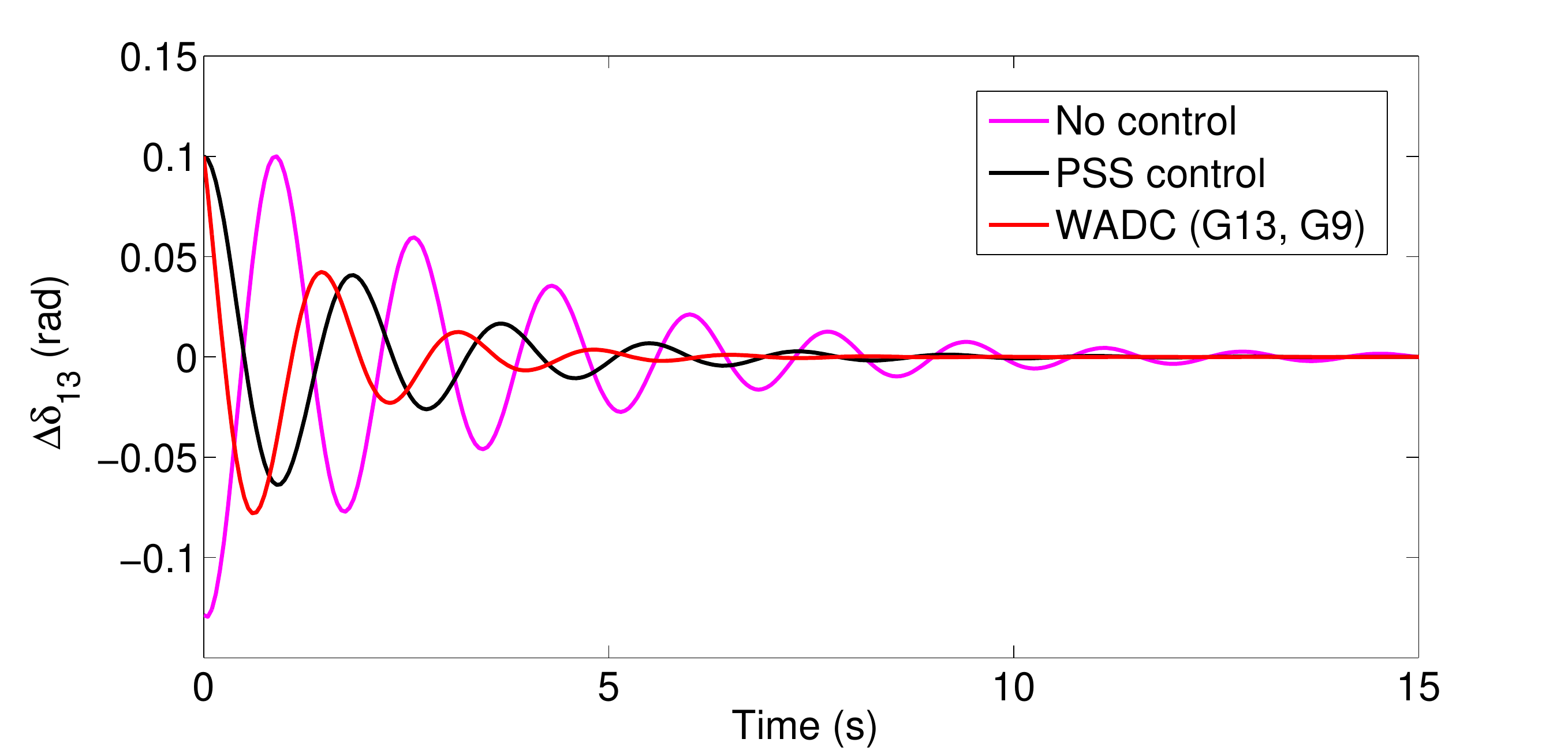}
}
\hfill
\subfloat[]{\includegraphics[width=3in]{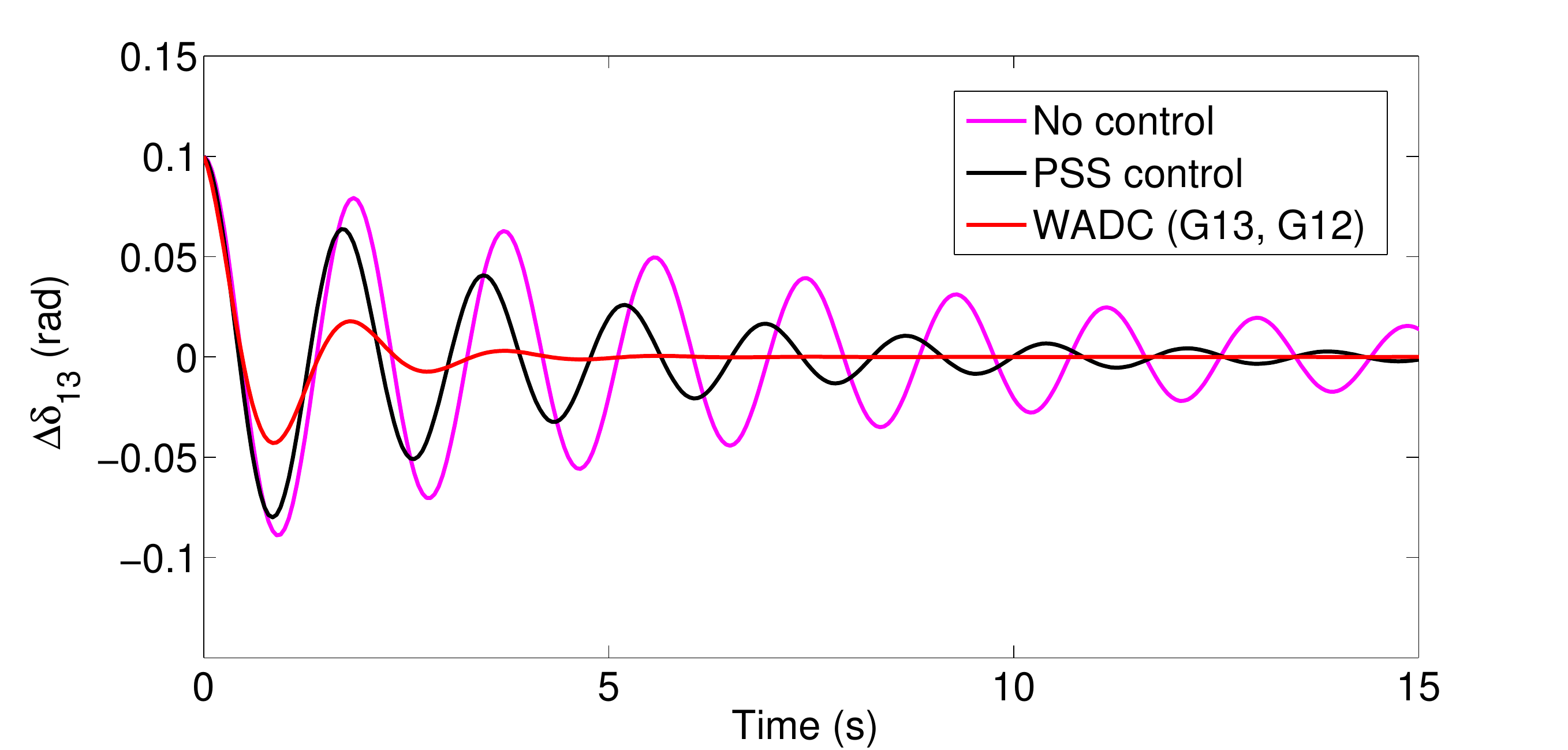}
}
\caption{Time-domain response of $\Delta\delta_{13}$ to the excitation of Mode 2. (a) Nominal condition. (b) Topology change condition.}\label{prepe}
\end{figure}

To further illustrate the effectiveness of the proposed WADC approach, 
Fig. \ref{prepe} presents a comparison of the time-domain response of $\Delta\delta_{13}$ to the excitation of Mode $2$ for different control scenarios and operating conditions. 
Mode $2$ is excited by setting $\bm{x_0}=\phi_{2_+}+\phi_{2_-}$, where $\bm{x_0}=[\Delta\bm{\delta_0},\Delta\bm{\omega_0}]^T$ is the initial condition of the system and $\phi_{2_+}$, $\phi_{2_-}$ are the right eigenvectors corresponding to the complex conjugate eigenvalues $\lambda_{2_+}$ and $\lambda_{2_-}$ of Mode $2$, respectively, similar to the approach adopted in \cite{Kundur94, Iravani13, Far09}. 
It can be seen that the inter-area oscillation associated with Mode $2$ always settles within $10s$ when the proposed WADC technique is performed. On the other hand, PSS control achieves the damping goal only without the network model change. 

\subsubsection{Performance Robustness}

Practical power systems undergo several disturbances and changes in operating conditions such as transmission line outages, short circuits, load trips, generator set-point shifts, etc. Therefore, the performance of wide-area damping control schemes should be tested over a wide range of operating conditions and various types of realistic contingencies. To this end, we assess the performance robustness of the proposed WADC approach, in comparison with the traditional PSS, for the following operating conditions:
\begin{enumerate}[(a)]
    \item Normal condition 
    \item Three-phase fault at Bus 18 at $t= 1s$   
    \item 20\% load trip at Bus 17 at $t= 1s$ 
\end{enumerate}

It should be noted that the three-phase fault at Bus $18$ is cleared in $3$ cycles (i.e., $0.05 s$) by tripping the line $18-49$. Simulations are conducted on 6th-order generator model, while $\sigma_i=5$ (see (\ref{swingrandom-2})). The goal of the proposed WADC algorithm is to damp all the critical inter-area modes, i.e., Mode $2$ and Mode $3$, 
under all operating conditions.
Table \ref{condition_results1} shows the estimated frequency, damping ratio and settling time of Mode $2$ under no control, PSS control and WADC with $\sigma_{d_{k}}=2, k\in\{2,3 \}$. Similar results can be obtained for Mode $3$.

It has been shown that the proposed WADC method can always damp effectively Mode $2$ (also Mode $3$) even though different kinds of disturbances and operating condition changes happen. The reason for this robust behavior is its model-free nature. After a contingency, the WADC algorithm simply employs the post-disturbance PMU measurements to estimate the inter-area mode features in real-time and thus adaptively modify the control signals $\bm{u}$ in accordance with the updated power network model. On the other hand, PSS provides the desired damping performance to Mode $2$ (i.e. $\zeta_{cl_2}\geq0.1$ and $t_{Scl_2}\leq10s$) 
only under the normal operating condition. Since its parameters cannot be re-configured online, an offline process is needed to respond to a disturbance and take into account the new inter-area mode properties. Note that the PSS is ineffective in suppressing Mode $3$ under any conditions, as its frequency lies beyond the limited PSS bandwidth. 


\begin{table}[!t]
\centering

  \caption{Estimated Properties of Mode 2 under Different  Operating Conditions and Control Methods}\label{condition_results1}
\setlength{\tabcolsep}{2.6pt} 
  \begin{tabular}{c c c c c}
\hhline{=====}
\multirow{3}{4em}{Operating Condition} & \multirow{3}{4em}{Control Method} &&&\\
&&Est. $f$ (Hz) &Est. $\zeta (\%)$ &Est. $t_{S}$ (s)\\
&&&&\\
  \hline
Normal&No control&0.587&8.220&13.157\Tstrut\\
          &PSS control&0.545&14.119&8.187\\
          &WADC (G13, G15)&0.594&16.376&6.453\Bstrut\\\hline
Three-phase fault&No control&0.556&4.759&24.045\Tstrut\\
           (Bus 18)&PSS control&0.597&8.175&13.002\\
           &WADC (G13, G15)&0.553&22.479&4.988\Bstrut\\\hline
20\% load trip&No control&0.622&7.182&14.207\Tstrut\\
           (Bus 17)&PSS control&0.650&9.215&10.586\\
           &WADC (G13, G15)&0.630&16.142&6.179\Bstrut\\\hline
\hhline{=====}
  \end{tabular}\label{prevspost}
 \end{table}
		
\subsection{WADC Operation in Large-Scale Power Systems}

The applicability of the proposed wide-area damping control algorithm 
is further validated in a larger system---the IEEE 145-bus, 50-generator, 2-area power system \cite{Vittal92}, which represents an approximated model of an actual power system and has been widely used for stability studies in the literature \cite{Reeve04, Ngamroo19}. In this benchmark system (Fig. 7), six machines, i.e., $\{G93, G104, G105, G106, G110, G111\}$ are represented by the 4th-order generator model and equipped with AVRs and PSSs, while $G93$ serves as the reference machine. The rest of the synchronous machines are modelled as classical 2nd-order generators. 
The power system data 
for the 145-bus network can be found in: https://github.com/zenili/Wide-Area-Damping-Control-2019.

Assuming PMUs are installed at all generator buses, rotor angles and frequencies are obtained from the emulated PMU data using a window size of $180s$ and a sampling rate of 30 Hz. Low-magnitude random load variations with $\sigma_i=0.1$ have been applied. 
A comparison between the true and the estimated inter-area mode fundamental properties is demonstrated in Tables \ref{freq_open_145}-\ref{settling_open_145}. All inter-area modes are estimated with relatively good accuracy given the system scale. The small sample size  also guarantees a near real-time operation. Similarly, as was in the case of 68-bus system, all inter-area modes are critical (i.e., $n(Cr)=5$) and require additional damping.   

The proposed wide-area damping control technique is applied to enhance the damping performance of all inter-area modes. Table \ref{closed_loop_results_inc_145} presents the WADC results, which clearly confirm the effectiveness of the WADC algorithm in damping all critical modes simultaneously while using only 7 
generators, i.e., 
$15\%$ of the total number of  generators. The accuracy and effectiveness achieved in the 145-bus system show that the proposed WADC method 
is applicable to large-scale power systems. 

\begin{figure*}[!ht]
\centering
\label{BIG_SYS}
\includegraphics[width=5.2in ,keepaspectratio=true,angle=0]{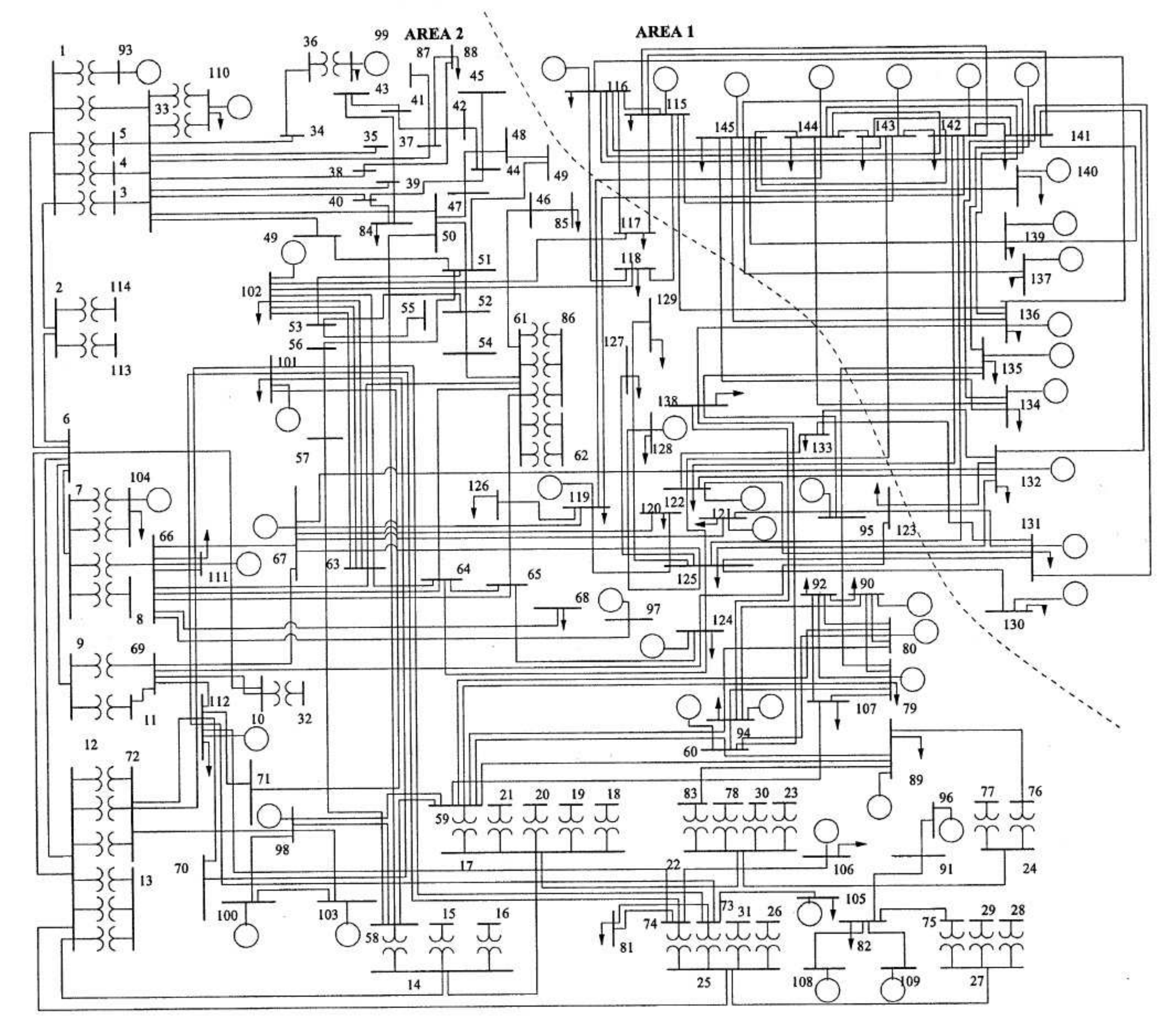}
\caption{145-bus, 50-generator, 2-area benchmark system \cite{Vittal92, Reeve04}.}
\end{figure*}

 \begin{table}[!t]
\centering
\caption{Actual and Estimated Inter-Area Mode Frequencies}\label{freq_open_145}
\begin{tabular}{c c c c}
\hhline{====}
\hline
Mode&Actual $f$ (Hz) & Estimated $f$ (Hz) &Error (\%) \Tstrut\Bstrut\\
  \hline
1&0.267&0.256&4.120\Tstrut\\
2&0.457&0.456&0.219\\
3&0.594&0.600&1.010\\
4&0.760&0.738&2.895\\
5&0.984&0.987&0.305\Bstrut\\
\hhline{====}
  \end{tabular}
 \end{table}

 \begin{table}[!t]
\centering
  \caption{Actual and Estimated Inter-Area Mode Damping Ratios}\label{damping_open_145}
  \begin{tabular}{c c c c}
\hhline{====}
\hline
Mode&Actual $\zeta$ (\%) & Estimated $\zeta$ (\%) &Error (\%) \Tstrut\Bstrut\\
  \hline
1&5.878&6.659&13.287\Tstrut\\
2&1.531&1.536&0.327\\
3&2.228&2.053&7.855\\
4&4.182&4.878&16.643\\
5&1.212&1.025&15.429\Bstrut\\
\hhline{====}
  \end{tabular}
 \end{table}

 \begin{table}[!t]
\centering
  \caption{Actual and Estimated Inter-Area Mode Settling  Times}\label{settling_open_145}
    \begin{tabular}{c c c c}
\hhline{====}
\hline
Mode&Actual $t_S$ (s) & Estimated $t_S$ (s) &Error (\%) \Tstrut\Bstrut\\
  \hline
1&40.500&37.328&7.832\Tstrut\\
2&90.901&90.891&0.011\\
3&48.086&51.690&7.495\\
4&20.004&17.669&11.673\\
5&53.386&62.933&17.883\Bstrut\\
\hhline{====}
  \end{tabular}
 \end{table}

\begin{table}[!t]
\centering
  \caption{Comparison Between the Open-Loop and the Closed-Loop Properties of the Critical Modes}\label{closed_loop_results_inc_145}
  \setlength{\tabcolsep}{0.85pt}
  \begin{tabular}{c c c c c c}
\hhline{======}
\hline
Mode&Est. $\zeta_{ol}$ (\%)&$\zeta_{cl}$ (\%)&Est. $t_{Sol}$ (s) &$t_{Scl}$ (s) &${\bm O_G}$\Tstrut\Bstrut\\
  \hline
  1&6.659&55.405&37.328&4.453&\textbf{\{G118, G119, G128,}\Tstrut\\
  2&1.536&27.212&90.891&5.140&\textbf{G131, G139, G140, G141\}}\\
  3&2.053&20.483&51.690&5.132\\
  4&4.878&11.879&17.669&7.342\\
  5&1.025&11.406&62.933&5.673&\Bstrut\\\hline
\hhline{======}
\end{tabular}
 \end{table}

\section{Conclusion}\label{5}

This paper presents a novel PMU-based wide-area damping control algorithm to mitigate the poorly damped inter-area modes in power grids. The proposed method is independent of 
the network model 
and does not need any information beyond the generator inertia and damping constants. 
Numerical simulations demonstrate that the proposed WADC algorithm can adequately damp all the critical inter-area modes 
using minimum control effort in near real-time. In addition, it has been shown that
the method is  robust against PMU measurement noise, can adapt to power network changes, and is able to maintain its functionality despite PMU loss. 
Based on the above, it is believed that the proposed WADC approach can be promisingly integrated into 
dynamic security assessment (DSA) tools for continuous inter-area mode monitoring and control under massive renewable energy sources uncertainties. Further studies will focus on applying the proposed WADC algorithm using FACTS devices.%
%




\appendix
\section{}\label{appendix}
The stationary covariance matrix $C_{\bm x \bm x}$ satisfies the well-known Lyapunov equation \cite{Gardiner09}:
\begin{equation}
\label{eq:lyapunov}
AC_{\bm x \bm x} + C_{\bm x \bm x}A^T = -BB^T
\end{equation}
Given the expressions of $A$, $B$ and $C_{\bm x \bm x}$, one can obtain:
\begin{equation}
\label{dyn_jacobi_der}
C_{\omega\omega}-M^{-1}(\frac{\partial \bm{P_{e}}}{\partial \bm{\delta}}C_{\delta\delta}+DC_{\omega\delta}) = 0
\end{equation}
which can be re-written in the form of (\ref{dyn_jacobi}). Note that $C_{\bm x \bm x}$ is practically unknown due to limited data and thus the sample covariance matrix $Q_{\bm x \bm x}$ is used in its place. It can be shown that $Q_{\bm x \bm x}$ can be estimated from PMU measurements \cite{Bialek17}.
\begin{IEEEbiography}[{\includegraphics[width=1in,height=1.25in, keepaspectratio]{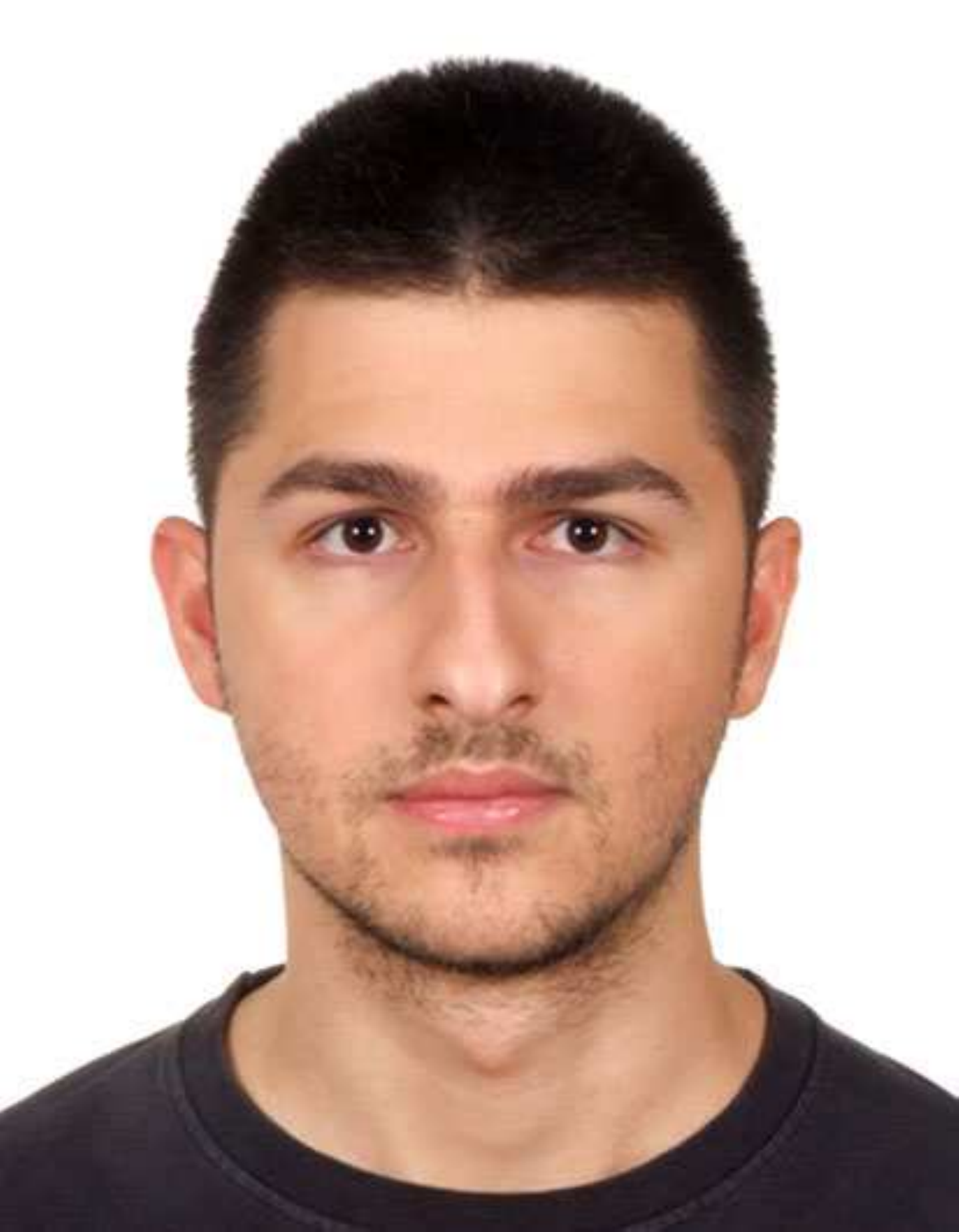}}]%
{Ilias Zenelis}
received a Diploma in Electrical and Computer Engineering from the National Technical University of Athens, Athens, Greece, in 2017. He is currently pursuing the Ph.D. degree in Electrical Engineering with the Electric Energy Systems Laboratory at McGill University, Montreal, Canada. His research interests include power system stability, electromechanical mode identification and wide-area damping control. 
\end{IEEEbiography}

\begin{IEEEbiography}[{\includegraphics[width=1.25in,height=1in,angle=-90, clip, keepaspectratio]{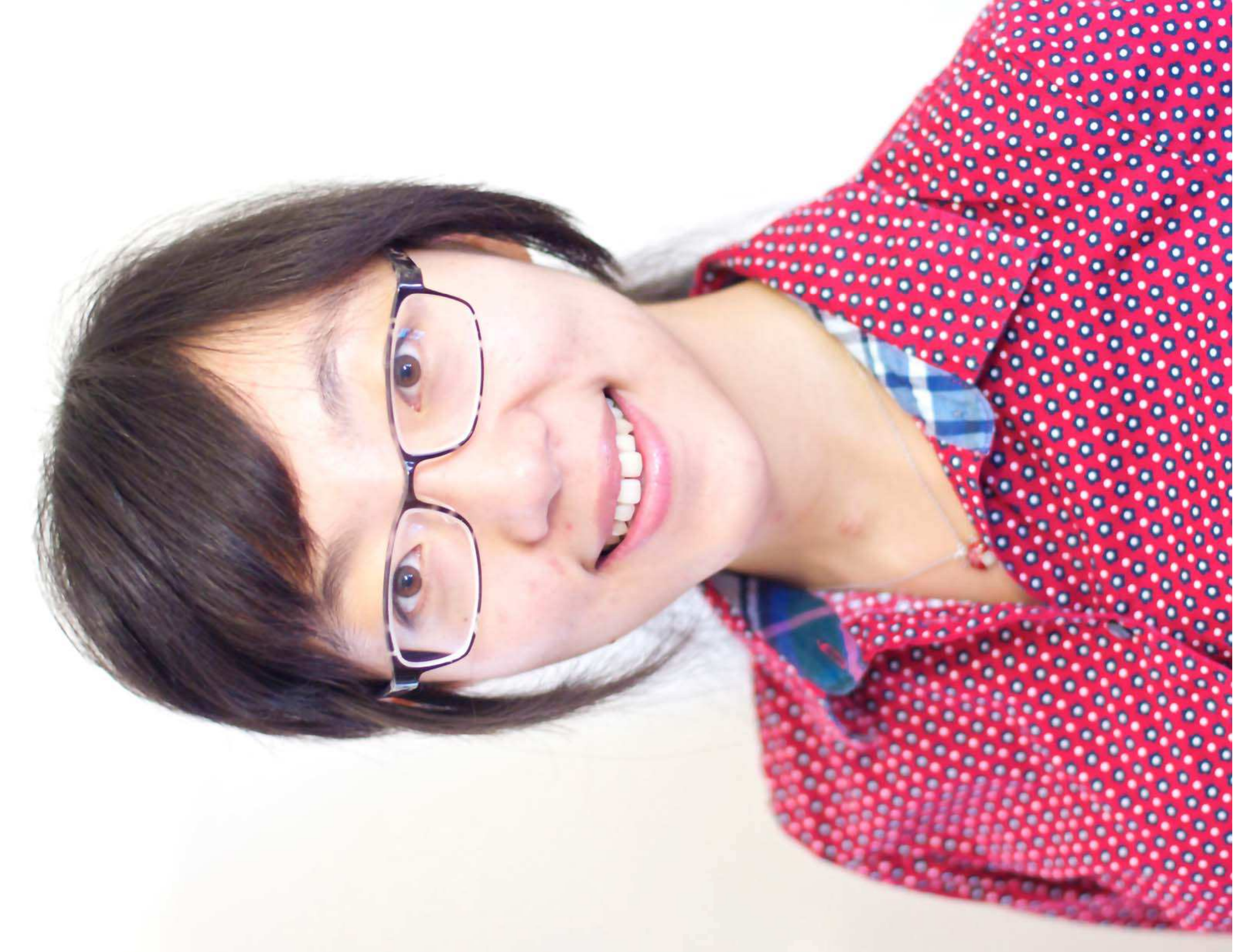}}]%
{Xiaozhe Wang}
is currently an Assistant Professor in the Department of Electrical and Computer Engineering at McGill University, Montreal, QC, Canada. She received the Ph.D. degree in the School of Electrical and Computer Engineering from Cornell University, Ithaca, NY, USA, in 2015. Her research interests are in the general areas of power system stability and control, uncertainty quantification in power system security and stability, and wide-area measurement system (WAMS)-based detection, estimation, and control. She is serving on the editorial boards of IEEE Transactions on Power Systems, Power Engineering Letters, and IET Generation, Transmission and Distribution.
\end{IEEEbiography}

\begin{IEEEbiography}[{\includegraphics[width=1in,height=1.25in, keepaspectratio]{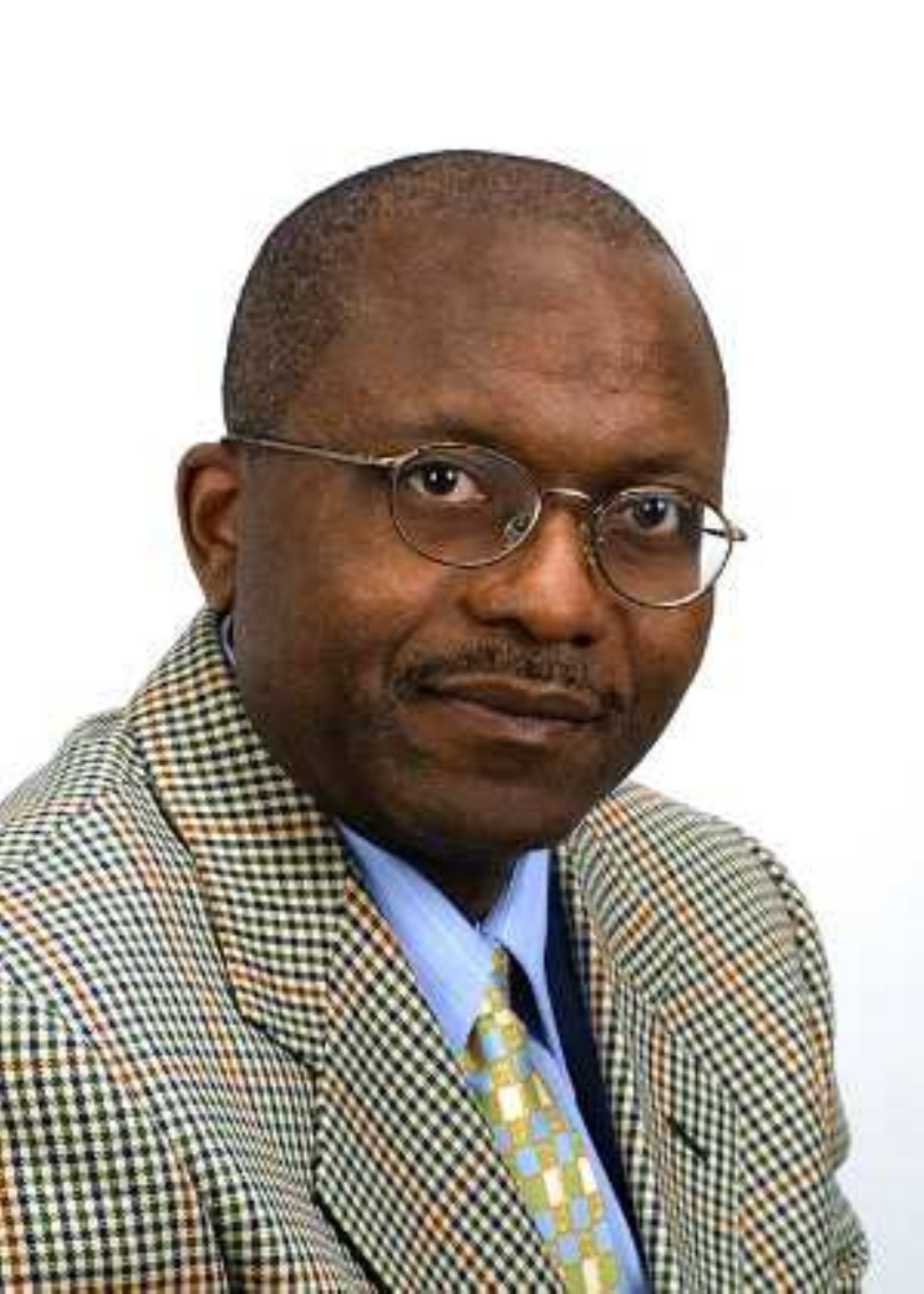}}]
{Innocent Kamwa}
obtained his B.S. and Ph.D. degrees in Electrical Engineering from Laval University, Qu\'{e}bec City in 1985 and 1989 respectively.  He has been a research scientist and registered professional engineer at Hydro-Quebec Research Institute since 1988, specializing in system dynamics, power grid control and electric machines. After leading System Automation and Control R\&D program for years he became Chief scientist for smart grid, Head of Power System and Mathematics, and Acting Scientific Director of IREQ in 2016. He currently heads the Power Systems Simulation and Evolution Division, overseeing the Hydro-Quebec Network Simulation Centre known worldwide. An Adjunct professor at Laval University and McGill University, Dr. Kamwa's Honors include four IEEE Power Engineering best paper prize awards, Fellow of IEEE and Fellow of the Canadian Academy of Engineering.  He is also the Recipient of the 2019 IEEE Charles Proteus Steinmetz Award and 2019 IEEE/PES Charles Concordia Power Systems Engineering Award.
\end{IEEEbiography}

\end{document}